\documentclass[12pt]{article}

\usepackage[a4paper, margin=2.5cm]{geometry}
\usepackage{moreverb}
\usepackage{booktabs}

\usepackage[latin1]{inputenc}
\usepackage{subfig,bm,todonotes, amsmath}

\usepackage{multirow}

\usepackage{graphicx}
\usepackage{lipsum}
\usepackage{amsthm, amsmath, amssymb, amsfonts,verbatim}

\usepackage{natbib}
\usepackage{graphicx,epstopdf}
\usepackage{algpseudocode}
\usepackage{multirow}
\usepackage{booktabs}
\usepackage{animate}
\usepackage{graphicx}
\usepackage{booktabs}
\usepackage{multirow}
\usepackage{longtable}
\usepackage{enumerate}
\usepackage{color,shortvrb}
\usetikzlibrary{calc,arrows,positioning,matrix,fit,intersections,shapes,shapes.misc,snakes}
\usepackage{float}

\usepackage{lscape}
\usepackage{url}
\usepackage[breaklinks]{hyperref}
\usepackage{breakurl}
\usepackage{authblk}

\usepackage{bm}
\usepackage{subfig}
\usepackage{listings}

\setcitestyle{numeric,open={(},close={)}}

\newcommand{\pkg}[1]{{\normalfont\fontseries{b}\selectfont #1}}
\let\proglang=\textsf
\let\code=\texttt

\begin{document}

\title{\textbf{meta4diag: Bayesian Bivariate Meta-analysis of Diagnostic Test Studies for Routine Practice}}

\author{J. Guo} 
\author{A. Riebler} 
\affil{Department of Mathematical Sciences, Norwegian University of Science and Technology, Trondheim, PO 7491, Norway.}


\maketitle


\begin{abstract}
This paper introduces the \proglang{R} package \pkg{meta4diag} for implementing Bayesian bivariate meta-analyses of diagnostic test studies.
Our package \pkg{meta4diag} is a purpose-built front end of the \proglang{R} package \pkg{INLA}. While \pkg{INLA} offers full Bayesian inference for the large set of latent Gaussian models using integrated nested Laplace approximations, \pkg{meta4diag} extracts the features needed for bivariate meta-analysis and presents them in an intuitive way. It allows the user a straightforward model-specification and offers user-specific prior distributions. Further, the newly proposed penalised complexity prior framework is supported, which builds on prior intuitions about the behaviours of the variance and correlation parameters. Accurate posterior marginal distributions for sensitivity and specificity as well as all hyperparameters, and covariates are directly obtained without Markov chain Monte Carlo sampling. Further, univariate estimates of interest, such as odds ratios, as well as the SROC curve and other common graphics are directly available for interpretation.  An interactive graphical user interface provides the user with the full functionality of the package without requiring any \proglang{R} programming.
The package is available through CRAN \url{https://cran.r-project.org/web/packages/meta4diag/} and its usage will be illustrated using three real data examples.
\end{abstract}

\newpage
\section[Introduction]{Introduction}
A meta-analysis summarises the results from multiple studies
with the purpose of finding a general trend across the studies. It
plays a central role in several scientific areas, such as medicine,
pharmacology, epidemiology, education, psychology, criminology and
ecology \citep{borenstein2011introduction}. A bivariate meta-analysis
of diagnostic test studies is a special type of meta-analysis that
summarises the results from separately performed diagnostic test
studies while keeping the two-dimensionality of the data
\citep{van2002advanced, reitsma2005bivariate}. Results of a diagnostic
test study are commonly provided as a two-by-two table, which is a
cross tabulation including four numbers: the number of patients tested
positive that are indeed diseased (according to some gold standard),
those tested positive that are not diseased, those tested negative
that are however diseased and finally those tested negative that are
indeed not diseased. Usually the table entries are referred to as true
positives (TP), false positives (FP), false negatives (FN) and true
negatives (TN), respectively. Those entries are used to compute pairs
of sensitivity and specificity indicating the quality of the
respective diagnostic test. The main goal of a bivariate meta-analysis
is to derive summary estimates of sensitivity and specificity from
several separately performed test studies. For this purpose pairs of
sensitivity and specificity are jointly analysed and the inherent
correlation between them is incorporated using a random effects
approach \citep{reitsma2005bivariate,chu2006bivariate}. Related
accuracy measures, such as likelihood ratios (LRs), which indicate the
discriminatory performance of positive and negative tests, LR$+$ and
LR$-$ respectively, can be also derived. Further, frequently used
estimates include diagnostics odds ratios (DORs) illustrating the
effectiveness of the test or risk differences which are related to the
Youden index \citep{altman1990practical, youden1950}.

\citet{reitsma2005bivariate} proposed to model logit sensitivity and logit specificity using a bivariate normal likelihood, whereby the mean vector itself is modelled using a bivariate normal distribution (normal-normal model). Our new package \pkg{meta4diag} follows the approach proposed by \citet{chu2006bivariate} and \citet{hamza2008meta} using an exact binomial likelihood (binomial-normal model), which has been shown to outperform the approximate normal likelihood in terms of bias, mean-squared error (MSE) and coverage. Furthermore, it does not require a continuity correction for zero cells in the two-by-two table \citep{harbord2007unification}.
Recently, \citet{chen2011bayesian} and \citet{kuss2014meta} proposed a
third alternative, the beta-binomial model, where sensitivity and
specificity are not modelled after the logit transformation but on the
original scale using a beta distribution. The inherent correlation is
then incorporated via copulas \citep{kuss2014meta}. In the absence of
covariates or in the case that all covariates affect both sensitivity
and specificity \citep{harbord2007unification}, the binomial-normal
model can be reparameterised into the hierarchical summary receiver
operating characteristic (HSROC) model
\citep{rutter2001hierarchical,harbord2007unification}. In contrast to
the binomial-normal model the HSROC model uses a scale parameter and an accuracy parameter, which are functions of sensitivity and specificity and defines an underlying hierarchical SROC (summary receiver operating characteristic) curve.

Different statistical software environments, such as \proglang{SAS} software \citep{SAS-STAT}, \proglang{Stata} \citep{stata} and \proglang{R} \citep{r-project}, have been used in the past ten years to conduct bivariate meta-analysis of diagnostic test studies. Within a frequentist setting the \proglang{SAS} \texttt{PROC MIXED} routine and \texttt{PROC NLMIXED} routine  can be used to fit the normal-normal and binomial-normal model, see for example \citet{van2002advanced,arends2008bivariate,hamza2009multivariate}. The \proglang{SAS} macro \pkg{METADAS} provides a user-friendly interface
for the binomial-normal model and the HSROC model
\citep{takwoingi2008metadas}. Within \proglang{Stata} the module
\pkg{metandi} fits the normal-normal model using an adaptive
quadrature \citep{harbord2009metandi}, while the module \pkg{mvmeta}
performs maximum likelihood estimation of multivariate random-effects
models using a Newton-Raphson procedure
\citep{white2009multivariate, gasparrini2012mvmeta}. The \proglang{R} package \pkg{mada}
\citep{mada, doebler2012}, a specialised version of \pkg{mvmeta}, is
specifically designed for the analysis of diagnostic accuracy. The
package provides both univariate modelling of  log odds ratios and
bivariate binomial-normal modelling of sensitivity and specificity. A
continuity correction is used for zero cells in the two-by-two tables.

Since the number of studies involved in a meta-analysis of diagnostic
tests commonly is small, often less than 20 studies, and data within
each two-by-two table can be sparse, the use of numerical algorithms
for maximising the likelihood of the above complex bivariate model
might be problematic and lead to non-convergence
\citep{paul2010bayesian}. Bayesian inference that introduces prior
information for the variance and correlation parameters in the
bivariate term is therefore attractive
\citep{harbord2011commentary}. Markov chain Monte Carlo (MCMC)
algorithms can be implemented through the generic frameworks
\proglang{WinBUGS} \citep{lunn2000winbugs}, \proglang{OpenBUGS}
\citep{lunn2009bugs} or \proglang{JAGS} \citep{plummer2003jags}. There
exist further specialised R-packages for analysing diagnostic test
studies in Bayesian setting, such as \pkg{bamdit} or \pkg{HSROC}
\citep{verde2011package,schiller2012hsroc}. Instead of modelling the
link-transformed sensitivity and specificity directly, the package
\pkg{bamdit} models the differences ($D_i$) and sums ($S_i$) of the
link-transformed sensitivity and specificity jointly. The quantities $D_i$ and $S_i$ are roughly
independent by using these
linear transformations, so that  \cite{verde2010meta} used a zero centered prior for the
correlation of $D_i$ and $S_i$ to represent vague prior
information. Consequently, \proglang{JAGS} is used for model estimation. In contrast, \pkg{HSROC} builds on the
HSROC model to jointly analyse sensitivity and specificity with and
without a gold standard reference test. Uniform priors on a restricted
interval are thereby assumed for
all the hyperparameters and model estimation is carried out using a
Gibbs sampler \citep[chapter 10]{chen2013}.
However, the use of Bayesian approaches is still limited in practice which might be partly caused by the fact that many applied scientists feel not comfortable with using MCMC sampling-based procedures \citep{harbord2011commentary}. Implementation needs to be performed carefully to ensure mixing and convergence. Furthermore, MCMC based methods are often time consuming, in particular, when interest lies in simulation studies which require several MCMC runs.

\citet{paul2010bayesian} proposed to perform full Bayesian inference using integrated nested Laplace approximations (INLA) which avoids MCMC entirely \citep{rue2009approximate}. The R-package \pkg{INLA}, see \url{www.r-inla.org},  implements Bayesian inference using INLA for the large set of latent Gaussian models. However, we understand that the range of options and the required knowledge of available features in \pkg{INLA} might be overwhelming for the applied user interested in only one specific model.
Here, we present a new \proglang{R} package \pkg{meta4diag} which is a purpose-built package defined on top of \pkg{INLA} extracting only the features needed for bivariate meta regression.
Our package \pkg{meta4diag} implements the binomial-normal model.  Model definition is straightforward, and output statistics and graphics of interest are directly available. Therefore, users do not need to know the structure of the general \pkg{INLA} output object.
%
Although its greatest strength, another criticism towards Bayesian inference is the choice of prior distributions.
Our package \pkg{meta4diag} allows the user to specify prior distributions for the hyperparameters using intuitive statements based on the recently proposed framework of penalised complexity (PC) priors \citep{2014arXiv1403.4630S}. Alternatively, standard prior distributions or user-specific prior distributions can be used. Our package is appealing for routine use and applicable without any deep knowledge of the programming language \proglang{R} via the integrated Graphical User Interface (GUI) offering roll-down menus and dialog boxes implemented using the \proglang{R} package \pkg{shiny} \citep{shiny}.

The rest of this paper is organised as follows. In Section~\ref{sec:model} we introduce the binomial-normal model and discuss its estimation within a Bayesian inference setting. Here, specific emphasis is given on the definition of prior distributions. Section~\ref{sec:upg} illustrates the functionality of the package \pkg{meta4diag}. Model output and available graphics are described based on the previously analysed \code{Telomerase} \citep{glas2003tumor}, \code{Scheidler} \citep{scheidler1997radiological} and \code{Catheter} \citep{chu2009bivariate} datasets. Further, the user-friendly GUI is presented.
Finally, a conclusion is given in Section~\ref{sec:conc}.

\section[Model]{Introducing the statistical framework}\label{sec:model}
\subsection[bmodel]{Binomial-normal model for bivariate meta-analysis}
In a bivariate meta-analysis, each study presents the number of true positives (TP), false positives (FP), true negatives (TN), and false negatives (FN). Let $\textrm{Se}=\textrm{TP}/(\textrm{TP}+\textrm{FN})$ denote the true positive rate (TPR) which is known as sensitivity and $\textrm{Sp}=\textrm{TN}/(\textrm{TN}+\textrm{FP})$ the true negative rate (TNR) which is known as specificity.
\citet{chu2006bivariate} proposed the following bivariate generalised linear mixed effects model to summarise
the results of several diagnostic studies, $i=1, \ldots, I$, by modelling sensitivity and specificity jointly:
\begin{equation}
\begin{aligned}
\label{eq1}
&\textrm{TP}_i | \textrm{Se}_i\sim \textrm{Binomial}(\textrm{TP}_i+\textrm{FN}_i,\textrm{Se}_i),\quad \textrm{logit}(\textrm{Se}_i)=\mu + \mathbf{U}_i\bm{\alpha}+\phi_i,\\
&\textrm{TN}_i | \textrm{Sp}_i\sim \textrm{Binomial}(\textrm{TN}_i+\textrm{FP}_i,\textrm{Sp}_i),\quad \textrm{logit}(\textrm{Sp}_i)=\nu + \mathbf{V}_i\bm{\beta}+\psi_i,\\
&\begin{pmatrix} \phi_i \\ \psi_i \end{pmatrix} \sim \mathcal{N}\left[ \begin{pmatrix} 0 \\ 0 \end{pmatrix},\begin{pmatrix} \sigma_{\phi}^2 & \rho\sigma_{\phi}\sigma_{\psi} \\ \rho\sigma_{\phi}\sigma_{\psi} & \sigma_{\psi}^2 \end{pmatrix}  \right].
\end{aligned}
\end{equation}
\noindent Here,  $\mu$, $\nu$ denote the intercepts for $\text{logit}(\text{Se}_i)$ and $\text{logit}(\text{Sp}_i)$, respectively, and $\mathbf{U}_i$, $\mathbf{V}_i$ study-level covariates vectors with corresponding coefficient parameters $\bm{\alpha}$ and $\bm{\beta}$. The covariance matrix of the random effects $\phi_i$ and $\psi_i$ is parameterised using between-study variances $\sigma_{\phi}^2$, $\sigma_{\psi}^2$ and correlation $\rho$.

The most-commonly-used logit link function can be replaced by other monotone link functions, such as the probit or the complementary log-log transformation.
We assume that both sensitivity and specificity are modelled with the same link function.
If desired, model \eqref{eq1} can easily be changed to model sensitivity and the false positive rate ($1-\text{Sp}$), or the false negative rate ($1-\text{Se}$) and specificity, or
$1-\text{Se}$ and $1-\text{Sp}$, instead of sensitivity and specificity, causing the corresponding change in parameter estimates. Different model
options are available through the argument \code{model.type} in the package \pkg{meta4diag}, see Section~\ref{sec:data1}.

\subsection{Specification of prior distributions}
\label{sec:prior}
\begin{figure}[h!]
	\centering
	\includegraphics[width=0.6\textwidth]{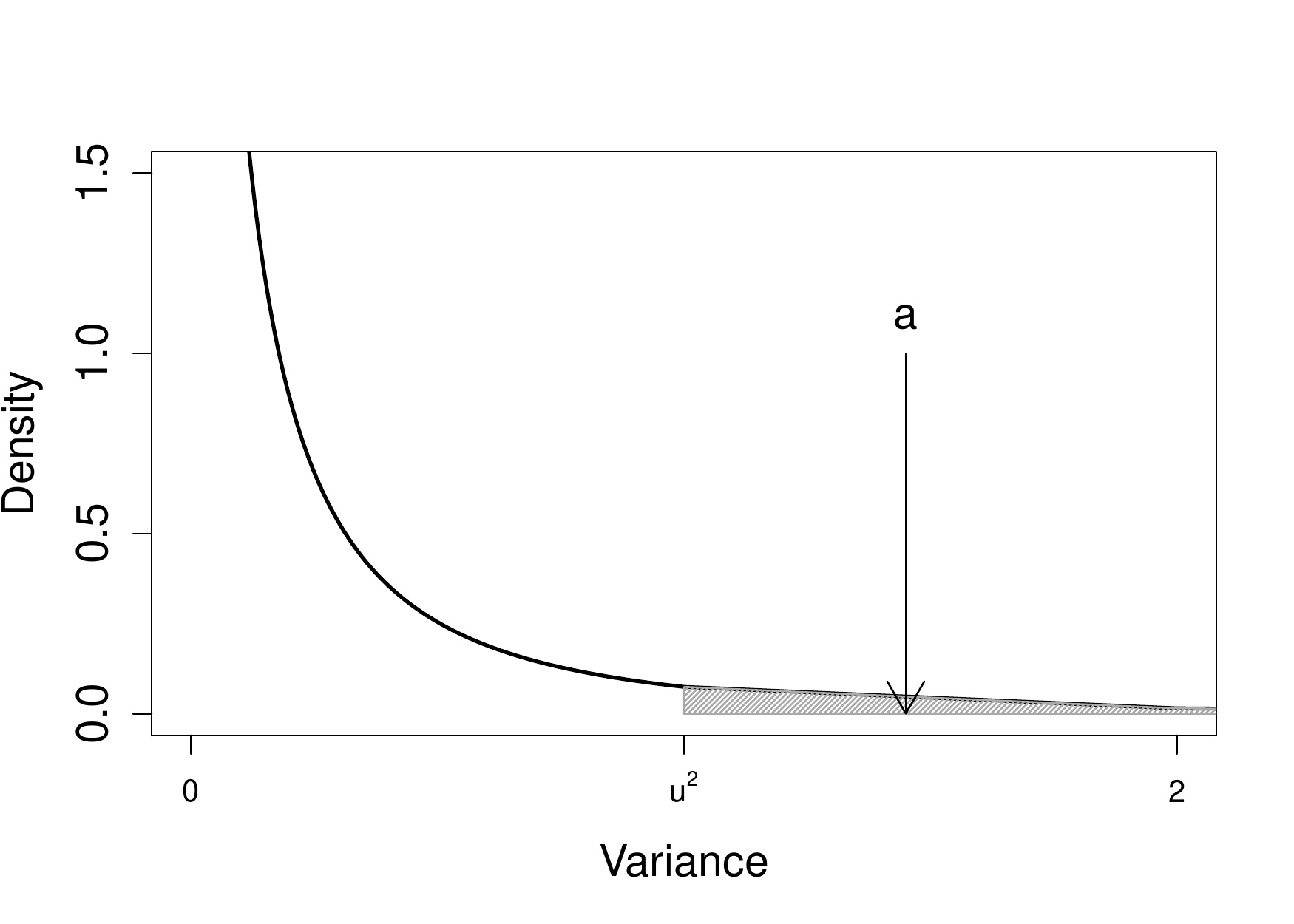}
	\caption{An example of the PC prior for the variance
          calibrated such that $P(\sigma>1)=0.05$. The black line is
          the prior density and the shaded area denotes the density
          weight $a=0.05$ when the standard deviation is larger than
          $u=1$. }
	\label{priorvariance}
\end{figure}

We specify prior distributions for all parameters, i.e.,~the three
hyperparameters $\sigma_{\phi}^2$, $\sigma_{\psi}^2$ and $\rho$, as
well as the fixed effects $\mu$, $\nu$, $\bm{\alpha}$ and
$\bm{\beta}$. Per default a normal prior with zero mean and large
variance is used for the fixed effects $\mu$, $\nu$, $\bm{\alpha}$ and
$\bm{\beta}$. The user is free to specify any prior distribution for
$\sigma_{\phi}^2$, $\sigma_{\psi}^2$ and $\rho$ including the newly
proposed penalised complexity (PC) priors, see
\citet{2014arXiv1403.4630S} for details. One of the four principles
underlying PC priors is Occam's razor. The idea is to see a certain
model component as a flexible extension of a base model (commonly a
simpler model) to which we
would like to reduce if not otherwise indicated by the data. Thinking
of a Gaussian random effect with mean zero and covariance matrix
$\sigma^2 \mathbf{I}$, the base model would be $\sigma^2=0$, i.e.,~the
absence of the effect. A PC prior puts maximum density mass at the base model
and decreasing mass with increasing distance away from the base model.
The PC prior for the variance components $\sigma_{\phi}^2$ or
$\sigma_{\psi}^2$ is discussed in
\citet[Section 2.3]{2014arXiv1403.4630S} and corresponds to an exponential prior
with parameter $\lambda$ for the standard deviation $\sigma_\phi$ or
$\sigma_\psi$, respectively. A simple choice to set $\lambda$ is to
provide $(u, a)$ such that $P(\sigma>u)=a$ leading to
$\lambda=-\log(u)/a$ with $u > 0$ and $0 < a < 1$. Figure~\ref{priorvariance} shows an example of
the PC prior for the variance. In practice, the PC prior for the
variance parameter in a diagnostic meta-analysis could be derived from
the belief of the interval that sensitivities or specificities lie
in. For example, choosing the contrast $\text{P}(\sigma>3)=0.05$
corresponds to believing that the sensitivities or specificities lie
in the interval $[0.5, 0.95]$ with probability $0.95$
\citep{wakefield2007disease}.

\begin{figure}[h!]
 \centering
 \subfloat[]{
    \includegraphics[width=0.33\textwidth]{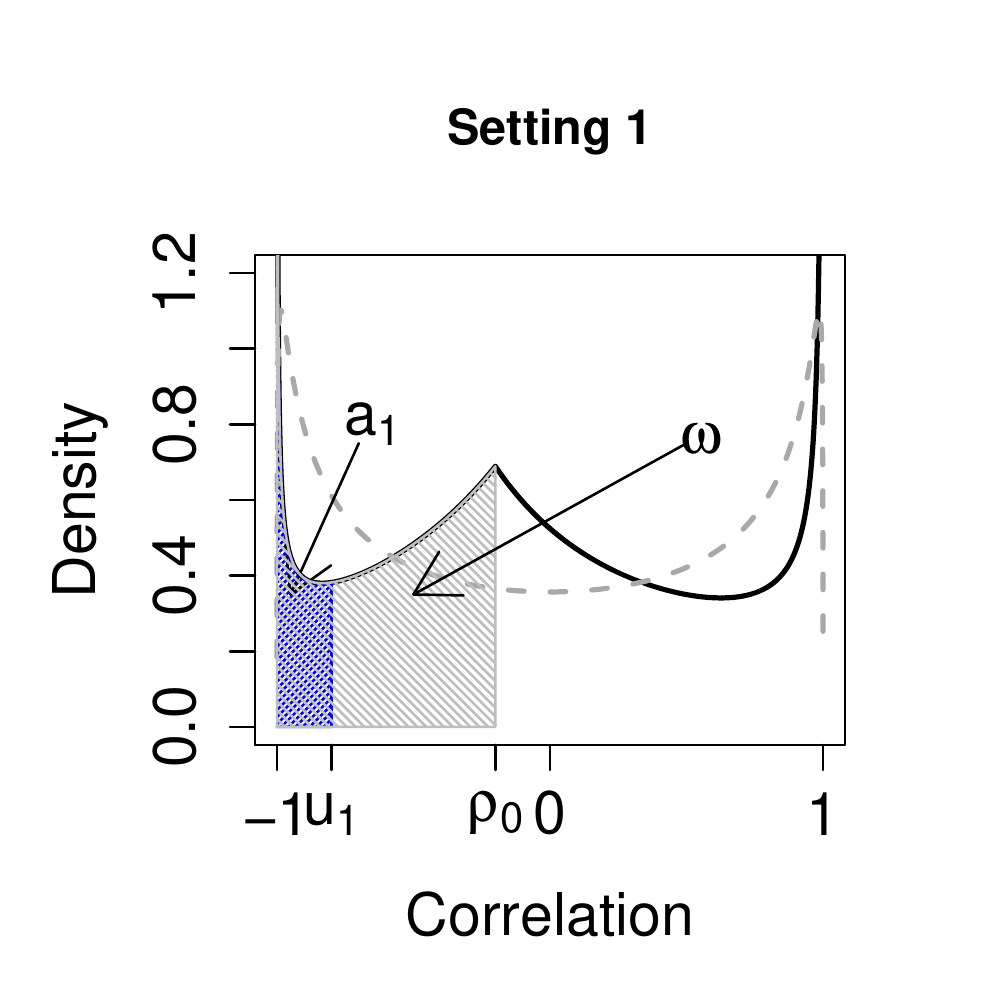}
    \label{a}
  }
  \subfloat[]{
    \includegraphics[width=0.33\textwidth]{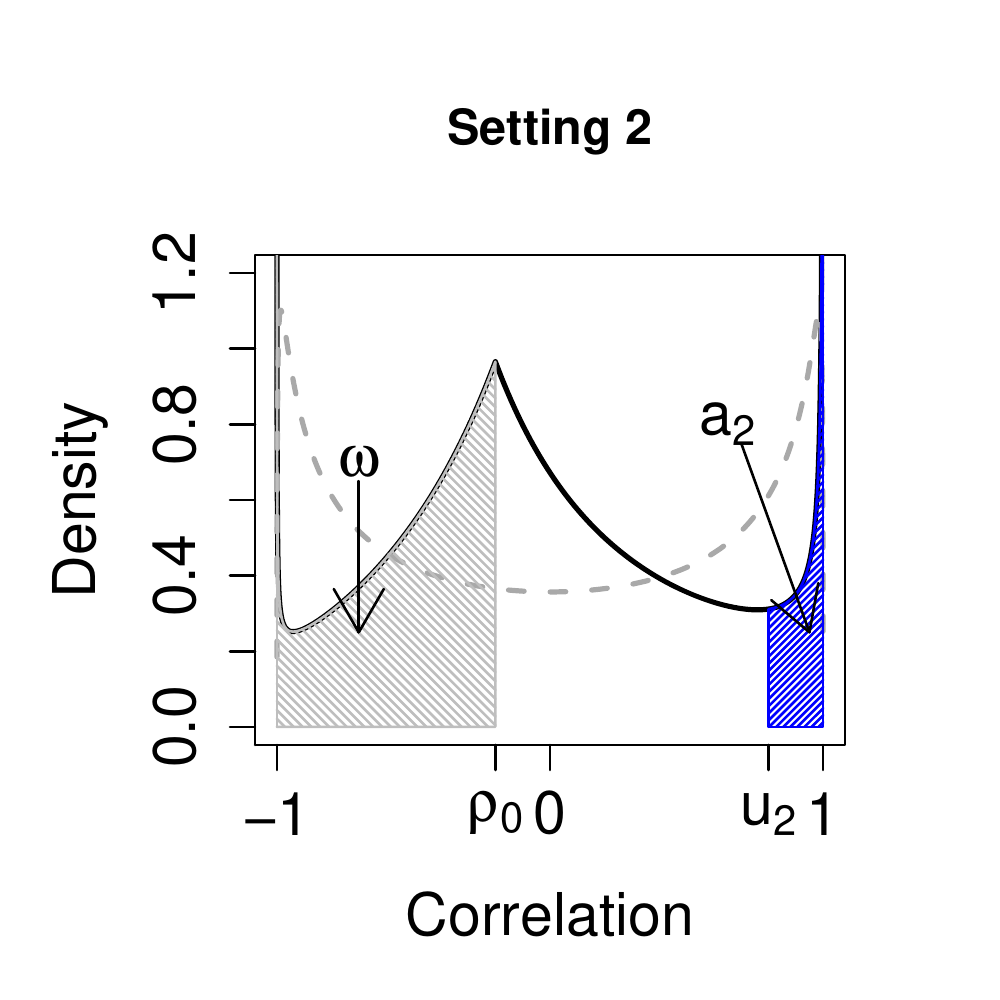}
    \label{b}
  }
  \subfloat[]{
    \includegraphics[width=0.33\textwidth]{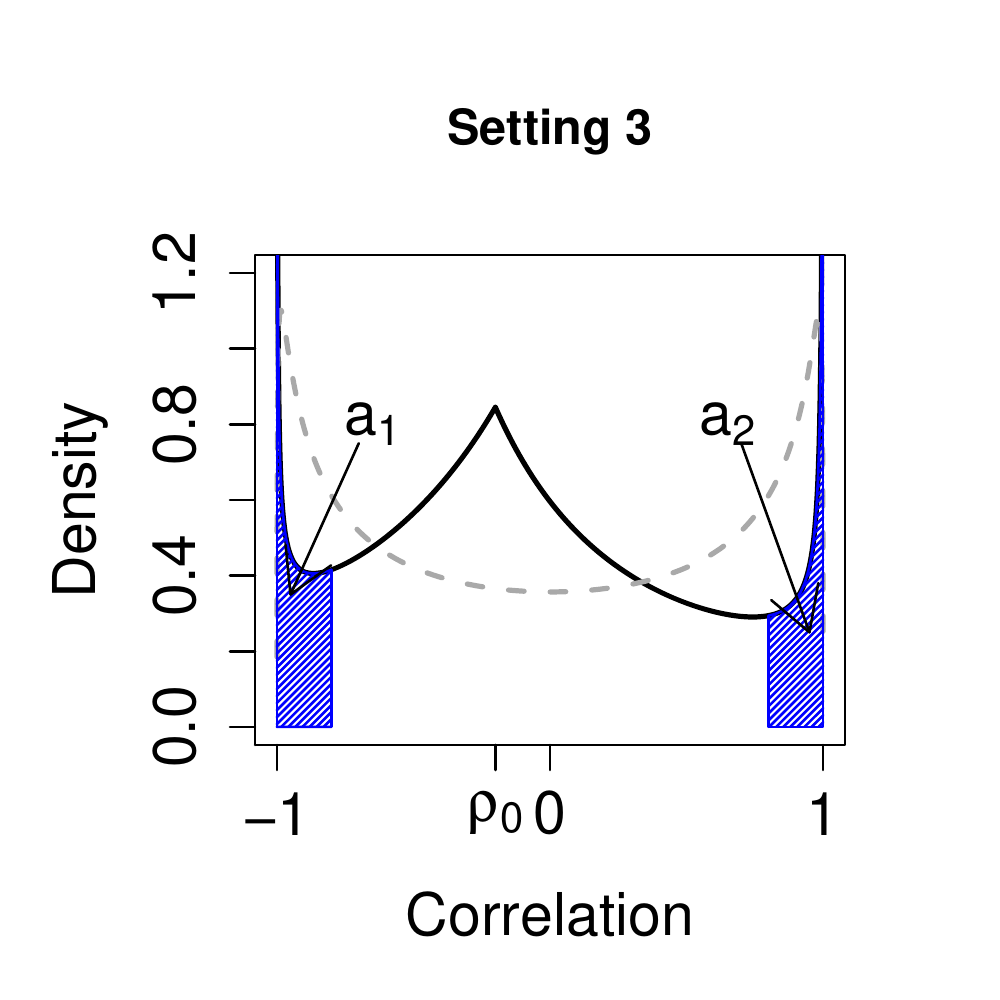}
    \label{c}
  }
	\caption{Illustration of potential PC priors for the correlation parameter $\rho$. The black solid line shows the PC prior and the dashed gray line shows the prior density proposed by \citet{paul2010bayesian}.
	In all plots we use $\rho_0=-0.2$. (a) Strategy 1: Prior
        derived using $P(\rho< -0.8| \rho_0 = -0.2)=0.1$ and $P(\rho<-0.2)=0.4$; (b) Strategy 2: Prior derived using $P(\rho>0.8| \rho_0 = -0.2)=0.1$ and $P(\rho<-0.2)=0.4$; (c) Strategy 3: Prior derived using $P(\rho< -0.8| \rho_0 = -0.2)=0.1$ and $P(\rho>0.8| \rho_0 = -0.2)=0.1$.}
	\label{priorcorrelation}
\end{figure}

For the correlation parameter $\rho$, \citet{harbord2011commentary}
proposed to use a stronger prior than the normal prior for the
Fisher's z-transformed correlation, which was used in
\citet{paul2010bayesian}. Motivated by the nature of diagnostic tests
he proposed to use a prior which is not centred around zero but
defined around some (negative) base value $\rho_0$ instead
\citep{reitsma2005bivariate}. Using the  PC prior framework the above
suggestions can be implemented  directly. \citet[Appendix
A.3]{2014arXiv1403.4630S} derives the PC-prior  for the correlation
parameter in an autoregressive model of first order assuming the base model being
defined at $\rho_0 = 0$ and identical statistical behaviour left and
right of $0$.  Although slightly tedious, this derivation can be
generalised to an  arbitrary $\rho_0$ and asymmetrical behaviour to
the left and  right of $\rho_0$ \citep{2015arXiv151206217G}. Within \code{meta4diag()} we offer
three strategies  to intuitively define a PC-prior for $\rho$ given an arbitrary value of $\rho_0$.
Similar as for the variance, probability contrasts are used to define
the prior intuitively.
\begin{description}
  \item[Strategy 1:] Specify the left tail behaviour and the probability mass on the left-hand side of $\rho_0$ by,
  		\begin{equation*}
		P(\rho<u_{1} | \rho_0)=a_{1} \quad\quad \text{and} \quad\quad P(\rho<\rho_{0})=\omega.
		\end{equation*}
		Here, $(\rho_0, \omega, u_1, a_1)$ are the hyperparameters needed to define the prior density.
  \item[Strategy 2:] Specify the right tail behaviour and the probability mass on the left-hand side of $\rho_0$ by,
  		\begin{equation*}
		P(\rho>u_{2}| \rho_0)=a_{2} \quad\quad \text{and} \quad\quad P(\rho<\rho_{0})=\omega.
		\end{equation*}
		Here, $(\rho_0, \omega, u_2, a_2)$ are the hyperparameters needed to define the prior density.
  \item[Strategy 3:] Specify left and right tail behaviours, by
  		\begin{equation*}
		P(\rho<u_{1}| \rho_0)=a_{1} \quad\quad \text{and} \quad\quad P(\rho>u_{2}| \rho_0)=a_{2}.
		\end{equation*}
		Here, $(\rho_0, u_1, a_1, u_2, a_2)$ are the hyperparameters needed to define the prior density.
\end{description}
Figure~\ref{priorcorrelation} shows examples of the PC prior for the
correlation using the three different strategies. The prior density
used in \citet{paul2010bayesian} is shown as the gray dashed lines for
comparison. The parameters for the strategies are motivated based on the
estimations results from \citet{menke2014bayesian}, who analysed $50$
independent bivariate meta-analyses which were selected randomly from
the literature within a Bayesian setting, and
\cite{diaz2015performance}, who reported frequentist estimates based
on a literature review of $61$  bivariate meta-analyses of diagnostic
accuracy published in 2010.
According to these two publications, the distribution of the
correlation seems asymmetric around zero.  We find that around half of
the correlation point estimates are negative, with a mode around
$-0.2$. Only a small proportion are larger than $0.4$ and values
larger than $0.8$ are rare.  Based on these findings, we choose three
differently  behaved PC priors that are all defined around $\rho_0 = -0.2$.

Defining the parameters of the prior distributions based on
probability contrasts seems very intuitive. As illustrated it is
straightforward to incorporate available prior knowledge into the prior
distributions, while still have the option to define vague priors
using less stringent probability contrasts.
Although we recommend to specify priors for the variance
and correlation components separately, our package also offers the option to use an inverse Wishart
distribution as a prior for the entire covariance matrix.

\section[Using package]{Using package \pkg{meta4diag}}\label{sec:upg}

\subsection{Package overview}
The \pkg{meta4diag} package provides functions for fitting bivariate meta-analyses within a full Bayesian setting as outlined in Section~\ref{sec:model}.
The package is available via CRAN \url{https://cran.r-project.org/web/packages/meta4diag/} and can be directly installed in \proglang{R} by typing

\code{\textit{R>} install.packages("meta4diag")}

(given a working internet connection and the appropriate access rights
on the computer). Within this paper we use package version 2.0.5 and \pkg{INLA} version 0.0-1466099681.
Of note, \pkg{meta4diag} requires \pkg{INLA} to be installed, which can be done using

\code{\textit{R>} install.packages("INLA", repos = "http://www.math.ntnu.no/inla/R/testing")}.

Once the package and its dependencies are installed all analyses presented throughout this work are reproducible.

The \pkg{meta4diag} package consists of one major function called \code{meta4diag()}. This function estimates the Bayesian bivariate regression model for diagnostic test studies, assuming each study provides TP, FP, TN and FN. Several studies can be grouped according to a categorical variable. Posterior estimates for parameters of the bivariate model as well as common plots and summary statistics are directly available. Inference is thereby performed using INLA, which provides accurate deterministic approximations to all model parameters and linear summary estimates. Based on the output of \code{meta4diag()} different plots of interest can be generated and also non-linear summary estimates, for example the diagnostics odds ratio (DOR), are available based on Monte-Carlo estimation, whereby iid samples are generated from the approximated posterior distribution using a built-in function of \pkg{INLA}.

The package includes three data sets which will be used in the following subsections to illustrate the functionality of \pkg{meta4diag}. The data sets differ in their structure and the availability of covariates. The first data set, called \code{Telomerase}, was presented by \citet{glas2003tumor} and consists of 10 diagnostic test studies. There is no covariate information available. The low number of studies involved makes this data set challenging when using maximum likelihood procedures, see for example \citet{riley2007bivariate, paul2010bayesian}. The second data set, called \code{Scheidler}, was presented in \cite{scheidler1997radiological} and combines three meta-analyses to compare the utility of three types of diagnostic imaging procedures to detect lymph node metastases in patients with cervical cancer. The third data set, called \code{Catheter}, consists of $33$ studies from a diagnostic accuracy analysis presented by \citet{chu2009bivariate} and provides disease prevalence as additional covariate.
\subsection{General data structure required}
\label{sec:data}
The first argument \code{data} in the function \code{meta4diag()} is the data set. It should be given as a data frame with a minimum of $4$ columns named \code{TP}, \code{FP}, \code{TN} and \code{FN}. If there is no column named \code{studynames} providing study names, the \code{meta4diag()} function will generate an additional column setting the study name indicators to \code{study$_1$}, $\cdots$, \code{study$_n$}, where $n$ is the number of studies in the meta-analysis. Further columns are considered to be covariates. The data set \code{Telomerase} can thus be defined using five columns, where the first column provides study name indicators and the remaining four provide values of TP, FP, TN and FN.

\begin{verbatim}
R> studynames <- c("Ito_1998", "Rahat_1998", "Kavaler_1998", "Yoshida_1997", 
+    "Ramakumar_1999", "Landman_1998", "Kinoshita_1997", 
+    "Gelmini_2000", "Cheng_2000", "Cassel_2001")
R> TP <- c(25, 17, 88, 16, 40, 38, 23, 27, 14, 37)
R> FP <- c(1, 3, 16, 3, 1, 6, 0, 2, 3, 22)
R> TN <- c(25, 11, 31, 80, 137, 24, 12, 18, 29, 7)
R> FN <- c(8, 4, 16, 10, 17, 9, 19, 6, 3, 7)
R> Telomerase <- data.frame(studynames = studynames,
+    TP = TP, FP = FP, TN = TN, FN = FN)
R> head(Telomerase)
\end{verbatim}
\begin{verbatim}
      studynames TP FP  TN FN
1       Ito_1998 25  1  25  8
2     Rahat_1998 17  3  11  4
3   Kavaler_1998 88 16  31 16
4   Yoshida_1997 16  3  80 10
5 Ramakumar_1999 40  1 137 17
6   Landman_1998 38  6  24  9
\end{verbatim}
\subsection[Example 1]{Analysing a standard meta-analysis without covariate information}
\label{sec:data1}

Here, we show how to analyse the \code{Telomerase} data set which represents a meta-analysis of studies that use the telomerase maker for the analysis of bladder cancer. To analyse the dataset, we first load the \pkg{INLA} and the \pkg{meta4diag} package in \proglang{R} using
\begin{verbatim}
R> library("INLA")
R> library("meta4diag")
\end{verbatim}
We then call the function \code{meta4diag()} as follows: 
\begin{verbatim}
R> res = meta4diag(data = Telomerase, model.type = 1, 
+    var.prior = "PC", var2.prior = "PC", cor.prior = "Normal", 
+    var.par = c(3, 0.05), cor.par = c(0, 5), 
+    link = "logit", nsample = 10000)
\end{verbatim}

The data set is transferred as the first argument followed by the
argument \code{model.type = 1}, saying that we would like to model
sensitivity and specificity jointly. Of note, the argument
\code{model.type} can be any integer from 1 to 4 depending on which
two accuracy measures are going to be modelled. When \code{model.type
  = 1}, sensitivity and specificity are modelled jointly. The
sensitivity and (1-specificity), (1-sensitivity) and specificity and
(1-sensitivity) and (1-specificity) will be jointly modelled when
\code{model.type = 2}, \code{model.type = 3} and \code{model.type =
  4}, respectively. The argument \code{var.prior} is a character
string to specify the prior distribution for the (transformed)
variance component of the first accuracy measure, i.e., here the
sensitivity. The options are \code{"PC"} for the PC prior,
\code{"Tnormal"} for the truncated normal prior, \code{"Hcauchy"} for
the half-Cauchy prior and \code{"Unif"} for the uniform prior, which
are all defined on the standard deviation scale. Alternatively
\code{"Invgamma"} for the inverse gamma prior or any user specified prior defined on the variance scale can be chosen. A user-specified prior for the variance is chosen by setting \code{var.prior = "Table"} and providing a 2-column data frame to \code{var.par}. The first column provides support points for the variance which should be in $[0, \infty]$, and the second column provides the corresponding prior density at these points. Of note, the usage of \code{"Table"} prior in \pkg{meta4diag} is different from that in \pkg{INLA}. While \pkg{INLA} requires the user to define the \code{"Table"} prior on the internal parameterisation of the hyperparameter, the user of \pkg{meta4diag} can work on the original scale. The argument \code{var2.prior} is a character string to specify the prior distribution for the second variance component. The options are the same as for the argument \code{var.prior}.

The argument \code{cor.prior} is a character string defining the prior
distribution for the (transformed) correlation parameter between the
two accuracy measures. The options are \code{"PC"} for the PC prior
defined on the correlation scale, \code{"Normal"} for the normal distribution defined on the Fisher's z-transformed correlation, \code{"Beta"} for the beta distribution defined on a suitable transformation, see documentation, and \code{"Table"} for an user specific prior defined on the correlation scale. The \code{"Table"} prior for the correlation should be provided as a 2-column data frame, where the first column provides suitable support points within $[-1, 1]$, and the second column provides the corresponding density mass of those points. Alternatively, if at least one of three arguments \code{var.prior}, \code{var2.prior} and \code{cor.prior} is set to \code{"Invwishart"}, an inverse Wishart distribution will be used for the covariance matrix ignoring any other prior definitions for the remaining arguments. The arguments \code{var.par}, \code{var2.par}, \code{cor.par} are numerical vectors specifying the hyperparameters for the priors for variance and correlation parameters. If the inverse Wishart prior is used the hyperparameters can be set in \code{wishart.par}. Prior definitions including parameterisations of the different options are given in the package documentation of \code{meta4diag()} or \code{makePriors()}. Of note, the arguments \code{var.prior}, \code{var2.prior} and \code{cor.prior} are not case sensitive, i.e. \code{var.prior="pc"} is valid if one uses it to indicate the PC prior for the first variance component.

Here, we use the logit link function by using \code{link = "logit"}. Alternative options are \code{"probit"} for the probit link and \code{"cloglog"} for the complementary log-log transformation. The argument \code{quantiles} requires a numerical vector with values in $[0, 1]$ defining which posterior quantiles should be returned. The default setting is \code{c(0.025, 0.5, 0.975)}, and these three quantiles will always be returned.  The argument \code{nsample} is an integer specifying the number of iid samples, generated from the approximated posterior distribution, which are used to compute any non-linear function of interest, such as DOR, LR$+$ or LR$-$.

To get summary information for all parameters of the model, we use the function \code{summary()}
\begin{verbatim}
R> summary(res)
\end{verbatim}
\begin{verbatim}
Time used: 
 Pre-processing    Running inla Post-processing           Total 
      0.1782410       0.1719451       0.2471750       0.5973611 

Fixed effects: 
    mean    sd 0.025quant 0.5quant 0.975quant
mu 1.179 0.198      0.788    1.178      1.577
nu 2.180 0.648      0.942    2.160      3.535

Model hyperpar: 
           mean    sd 0.025quant 0.5quant 0.975quant
var_phi   0.244 0.179      0.050    0.195      0.718
var_psi   3.647 2.070      1.144    3.137      9.071
cor      -0.819 0.200     -0.992   -0.888     -0.246

-------------------
          mean    sd 0.025quant 0.5quant 0.975quant
mean(Se) 0.763 0.030      0.703    0.765      0.818
mean(Sp) 0.887 0.052      0.762    0.896      0.963

-------------------
Correlation between mu and nu is -0.5504.
Marginal log-likelihood: -65.0459
Variable names for marginal plotting: 
      mu, nu, var1, var2, rho
\end{verbatim}

Here, also the time needed to fit the model as well as the estimated correlation between the two linear predictors, here $\mu$
and $\nu$,
is shown. This correlation is different from the hyperparameter correlation provided in \code{cor}, which corresponds to $\hat{\rho}$, i.e the posterior correlation between random effects.

\begin{figure}[h!]
\centering
\includegraphics{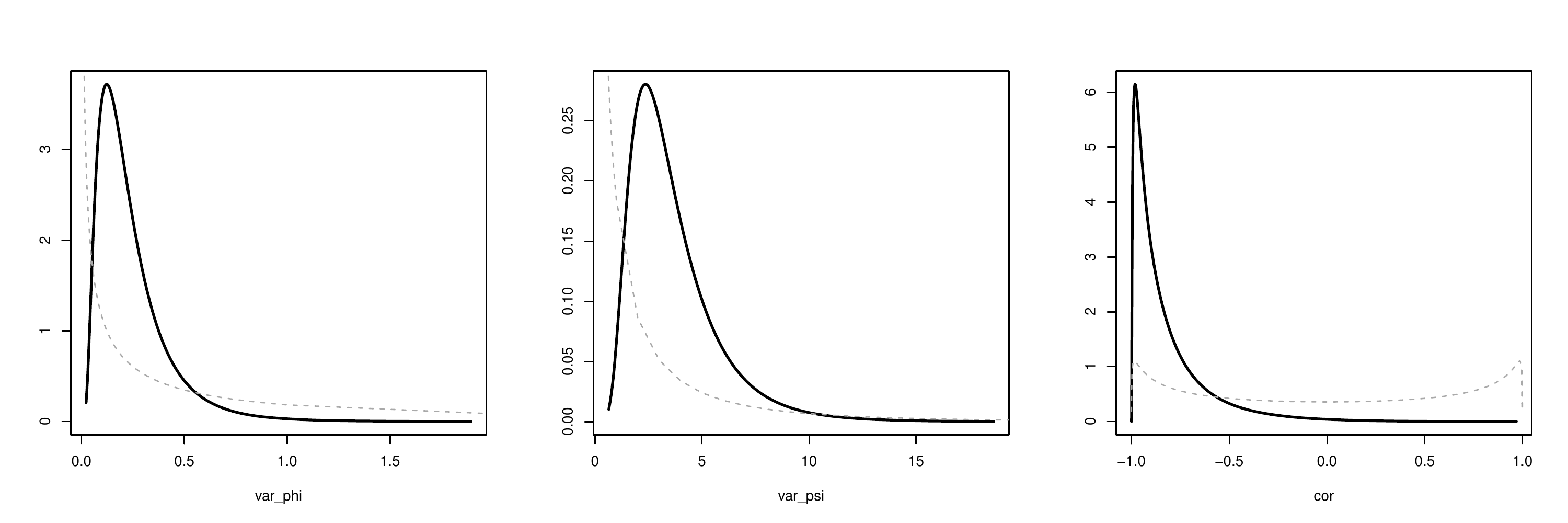}
\caption{Posterior marginals (black solid line) of $\sigma^2_{\phi}$, $\sigma^2_{\psi}$ and $\rho$ for the \code{Telomerase} data together with the prior distributions (gray dashed line).\label{fig:mp}}
\end{figure}

To plot the posterior marginal distribution of $\sigma^2_{\phi}$, say, we call the function \code{plot()} with argument \code{var.type = "var1"}. When defining separate prior distributions for the variance and correlation parameters and setting \code{overlay.prior = TRUE} the prior distribution is shown in the same device.
The posterior marginal distributions of $\sigma^2_{\phi}$ and
$\sigma^2_{\psi}$ together with their corresponding prior distribution
are shown in Figure~\ref{fig:mp}. Valid values of \code{var.type} are
the names of the fixed effects (i.e., \code{"mu"} and \code{"nu"} for
this dataset), \code{"var1"}, \code{"var2"} or \code{"rho"}. The
argument \code{save} can be set to \code{FALSE} (default) to
indicate that the resulting figures are not saved on the computer, or
to a file name, (e.g., \code{"posterior\_v1.pdf"}), to indicate that the
plot is saved as \code{"./meta4diagPlot/posterior\_v1.pdf"}, where
\code{"./"} denotes the current working directory and the directory
\code{meta4diagPlot} is created
automatically if it does not exist. Alternatively, the argument
\code{save} can be set to \code{TRUE} to indicate that the plot is
saved in the directory \code{meta4diagPlot} whereby the name is
chosen according to \code{var.type}.
Many standard \proglang{R}
plotting arguments, such as \code{xlab}, \code{ylab}, \code{xlim},
\code{ylim} and \code{col}, can also be set in the \code{plot()}
function.
\begin{verbatim}
R> par(mfrow = c(1, 3))
R> plot(res, var.type = "var1", overlay.prior = T, lwd = 2, save = F)
R> plot(res, var.type = "var2", overlay.prior = T, lwd = 2, save = F)
R> plot(res, var.type = "rho", overlay.prior = T, lwd = 2, save = F)
\end{verbatim}
\setkeys{Gin}{width=\textwidth}

To get descriptive statistics of study-specific accuracy measures of interest, such as positive or negative likelihood ratios LR$+$ or LR$-$, or the diagnostic odds ratio DOR, we call the function \code{fitted()}. The argument \code{accuracy.type} requires a single character string specifying the statistics of interest. Possible options are besides other \code{"sens"} (default), \code{"spec"},\code{"TPR"},\code{"TNR"},\code{"FPR"},\code{"FNR"}, \code{"LRpos"}, \code{"LRneg"}, \code{"RD"}, \code{"DOR"}, \code{"LLRpos"}, \code{"LLRneg"} and \code{"LDOR"} .
\begin{verbatim}
R> fitted(res, accuracy.type = "TPR")
\end{verbatim}
\begin{verbatim}
Diagnostic accuracies true positive rate (sensitivity): 
                mean    sd 0.025quant 0.5quant 0.975quant
Ito_1998       0.740 0.049      0.634    0.743      0.830
Rahat_1998     0.792 0.047      0.689    0.795      0.876
Kavaler_1998   0.827 0.030      0.766    0.828      0.883
Yoshida_1997   0.692 0.060      0.553    0.699      0.789
Ramakumar_1999 0.688 0.048      0.586    0.690      0.777
Landman_1998   0.794 0.039      0.712    0.796      0.865
Kinoshita_1997 0.623 0.068      0.479    0.627      0.742
Gelmini_2000   0.779 0.045      0.685    0.780      0.862
Cheng_2000     0.770 0.050      0.663    0.772      0.864
Cassel_2001    0.852 0.039      0.765    0.855      0.918
\end{verbatim}
\begin{verbatim}
R> fitted(res, accuracy.type = "DOR")
\end{verbatim}
\begin{verbatim}
Diagnostic accuracies diagnostic odds ratio (DOR): 
                  mean       sd 0.025quant 0.5quant 0.975quant
Ito_1998        68.499   61.320     13.572   51.707    225.479
Rahat_1998      18.770   12.137      4.819   15.924     50.201
Kavaler_1998    10.234    3.715      4.817    9.613     19.309
Yoshida_1997    69.950   41.273     20.087   60.082    174.630
Ramakumar_1999 203.266  186.636     45.031  149.894    692.741
Landman_1998    17.985    8.598      6.745   16.256     39.713
Kinoshita_1997 634.874 5401.401     11.485  130.997   3746.458
Gelmini_2000    34.573   24.859      8.747   28.289     99.020
Cheng_2000      35.889   21.476     10.840   30.989     89.925
Cassel_2001      2.683    1.354      0.885    2.410      5.991
\end{verbatim}

\begin{figure}[h!]
\centering
\includegraphics{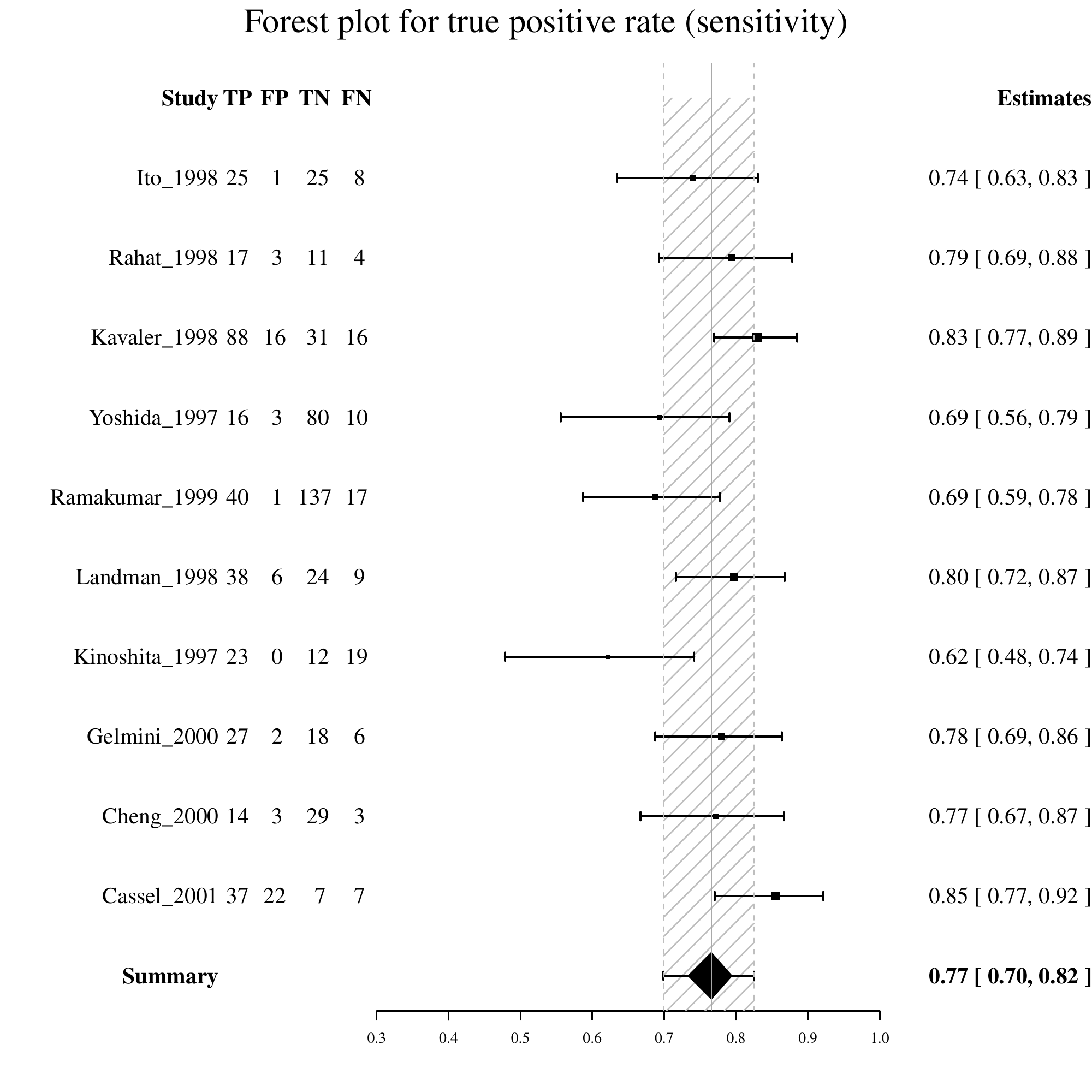}
\caption{Forest plot of the true positive rate (sensitivity) for the \code{Telomerase} data. Study names, the given observations (values of TP, FP, TN and FN) as well as model-based mean estimates within $95\%$ credible intervals are shown. At the bottom a summary estimate combining all studies is provided. The size of the study specified estimate points ($\blacksquare$) is proportional to the length of the corresponding credible intervals, the shorter the interval length the bigger the point and vice versa.}
\label{fig:forestplot}
\end{figure}

A commonly used graphic to illustrate the results of a meta-analysis is the so-called forest plot \citep{lewis-clarke-2001}. Figure~\ref{fig:forestplot} shows the forest plot including $95\%$ credible intervals for the telomerase data set as obtained using the \code{forest()} function.
\begin{verbatim}
R> forest(res, accuracy.type = "sens", est.type = "mean", cut=c(0.4, 1),
+    nameShow = T, ciShow = T, dataShow = "center", 
+    text.cex = 1.5, arrow.lwd = 1.5)
\end{verbatim}
\setkeys{Gin}{width=\textwidth}

The arguments \code{nameShow}, \code{dataShow}, \code{estShow} require a logical value indicating whether the study names, the given observations (values of TP, FP, TN and FN) and values of credible intervals are displayed as texts in the forest plot, respectively. The corresponding texts are right aligned when the arguments are set to be \code{TRUE}. They could also be \code{"left"}, \code{"right"} or \code{"center"} specifying the different alignments. The argument \code{accuracy.type} is defined as in the \code{fitted()} function. The argument \code{est.type} requires a character string specifying the summary estimate to be used. The options are \code{"mean"} (default) and \code{"median"}. The arguments \code{text.cex} specifies the text size, while \code{arrow.lwd} specifies the line width for the credible lines.

\begin{figure}[h!]
\centering
\includegraphics{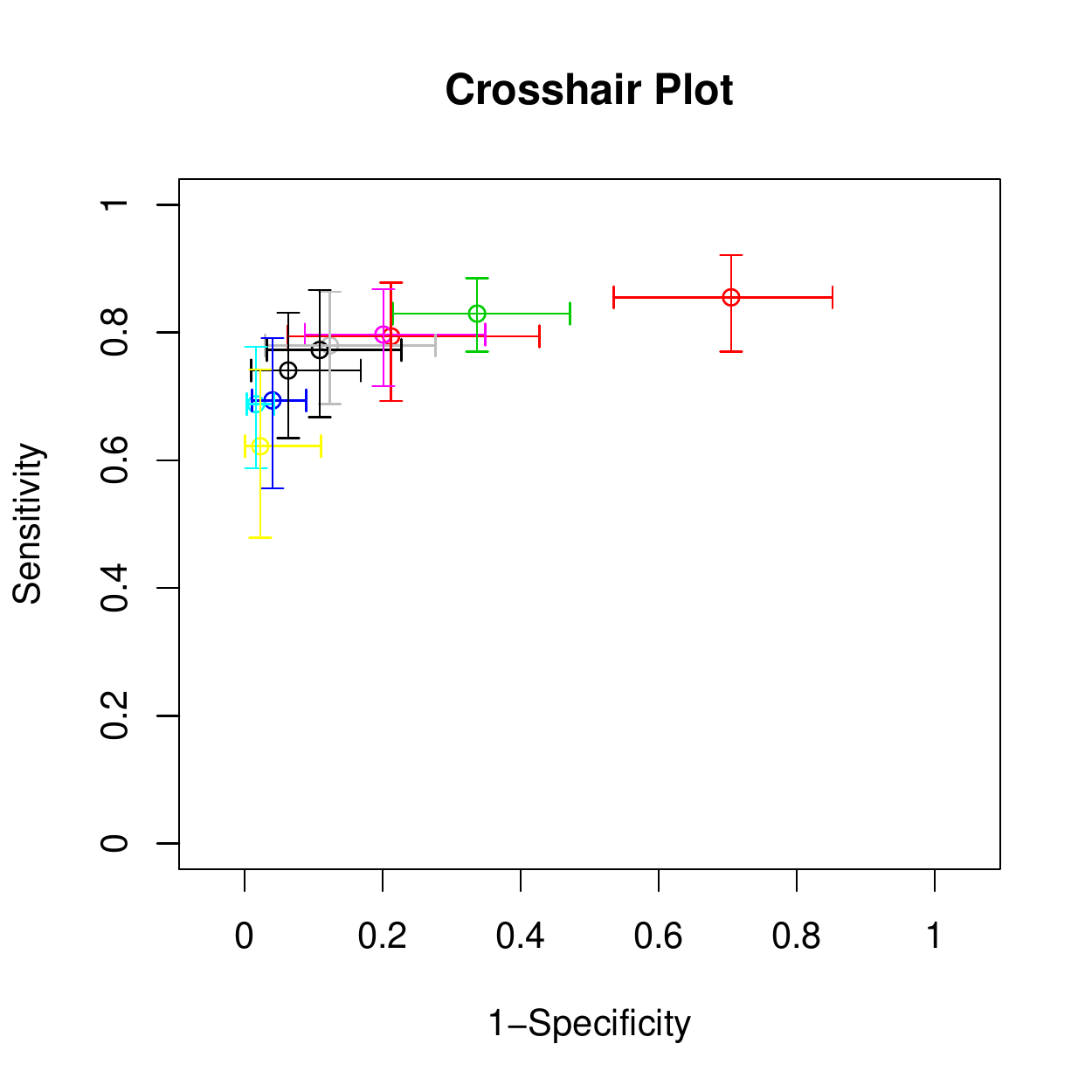}
\caption{Crosshair plot for the \code{Telomerase} data set. Shown are the posterior means for each study as the summary points together with paired lines showing the corresponding $95\%$ credible intervals for
sensitivity and (1-specificity). Colors are randomly chosen. \label{fig:cross}}
\end{figure}

\begin{figure}[h!]
\centering
\includegraphics{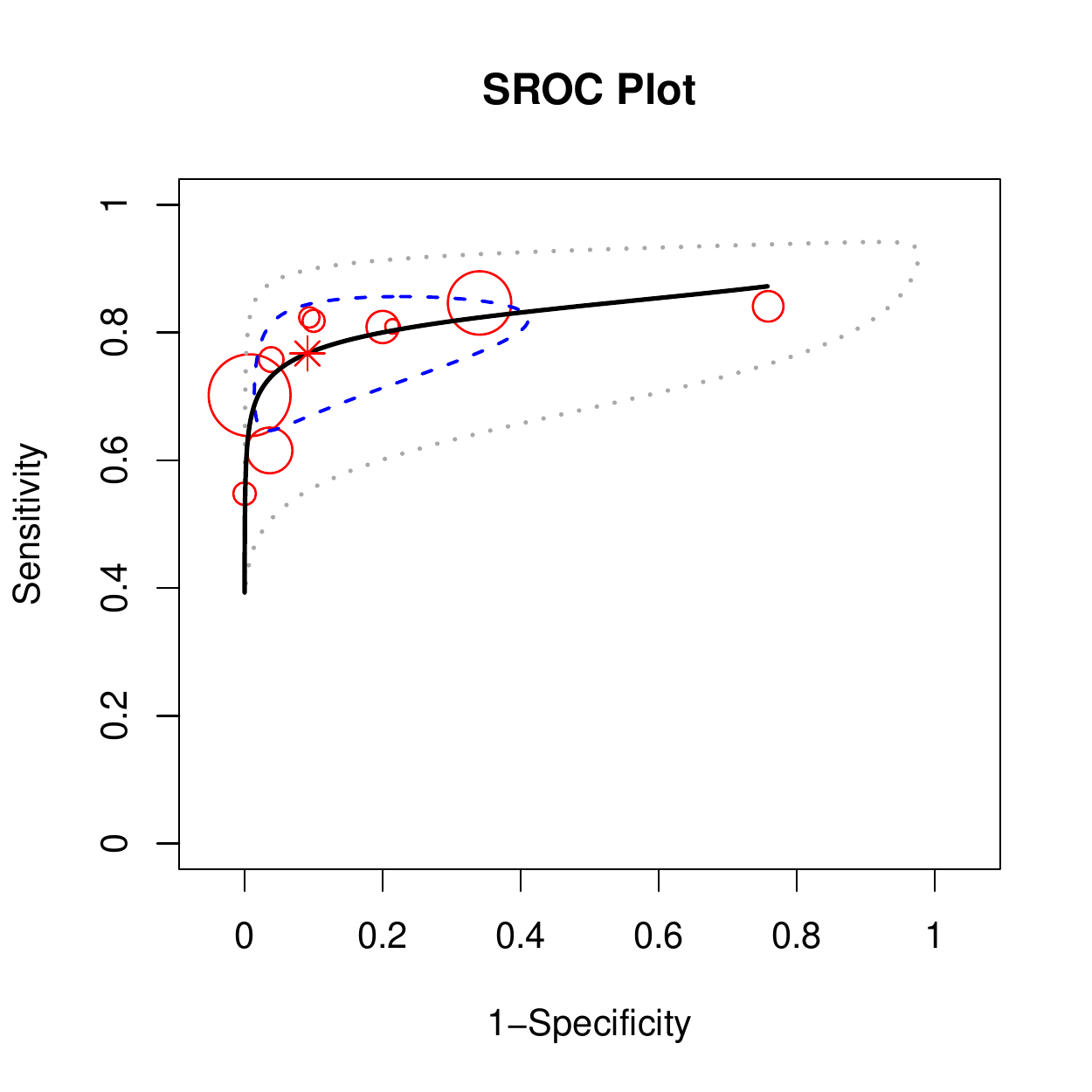}
\caption{SROC plot for the \code{Telomerase} data set. Each bubble represents one study and indicates its observed sensitivity and specificity. The size of the bubble is proportional to the number of individuals in this study. The solid black line is the SROC line. The star point represents the summary point, the dashed blue line is the $95\%$ credible region and the dashed gray line is the $95\%$ prediction region.\label{fig:sroc}}
\end{figure}

The two functions \code{crosshair()} and \code{SROC()} are available to study the result in the ROC space with sensitivity on the y-axis and 1-specificity on the x-axis. Figure~\ref{fig:cross} shows a crosshair plot displaying the individual studies in ROC space with paired confidence intervals representing sensitivity and specificity \citep{phillips-etal-2010}. Figure~\ref{fig:sroc} shows a summary receiver operating characteristic curve (SROC) which is only available when no separate covariates are included for the two model components, here sensitivity and specificity, as only then the bivariate meta-regression approach is equivalent to the HSROC approach \citep{rutter2001hierarchical}. The corresponding commands are
\begin{verbatim}
R> crosshair(res, est.type = "mean", col = c(1:10))
\end{verbatim}
\setkeys{Gin}{width=0.8\textwidth}

\begin{verbatim}
R> SROC(res, est.type = "mean", sroc.type = 1, 
+    dataShow = "o", crShow = T, prShow = T)
\end{verbatim}
\setkeys{Gin}{width=0.8\textwidth}

The argument \code{dataShow} specifies whether the original data are
shown. The argument \code{crShow}  and \code{prShow} are Boolean and
indicate whether a credible region or prediction region, respectively,
is shown. The argument \code{sroc.type} takes an integer value from 1
to 5. When \code{sroc.type=1}, the function used to define the SROC
line corresponds to `` The regression line 1" in
\citet{arends2008bivariate, chappell2009summary}. The  values \code{sroc.type=2},
\code{sroc.type=3}, \code{sroc.type=4} and \code{sroc.type=5}
correspond to ``The major axis method", ``The Moses and Littenberg's regression line",  ``The regression line 2" and ``The Rutter and Gatsonis's SROC curve", respectively.
Different choices  may result in different SROC lines when the correlation for sensitivity and specificity is positive. We refer to \citet{chappell2009summary} for more details and a comparison of the different formulations.

\subsection[Example 2]{Incorporating additional sub-data stratification} 
\label{sec:data2}
The \code{Scheidler} data set contains the results of a meta-analysis conducted by \citet{scheidler1997radiological} to compare the utility of three types of diagnostic imaging, lymphangiography (LAG), computed tomography (CT) and magnetic resonance (MRI), to detect lymph node metastases in patients with cervical cancer. The dataset consists of a total of $44$ studies: the first $17$ for CT, the following $17$ for LAG and the last $10$ for MRI. The \code{Scheidler} data set is provided in the package as a data frame with length $44$. It has a special column named ``modality" that specifies to which imaging technology, namely CT, LAG or MRI, each study belongs to. The first five lines of the data set are given as
\begin{verbatim}
R> data("Scheidler", package = "meta4diag")
R> head(Scheidler)
\end{verbatim}
\begin{verbatim}
           studynames modality TP FP FN TN
1       Grumbine_1981       CT  0  1  6 17
2          Walsh_1981       CT 12  3  3  7
3        Brenner_1982       CT  4  1  2 13
4     Villasanta_1983       CT 10  4  3 25
5 vanEngelshoven_1984       CT  3  1  4 12
6          Bandy_1985       CT  9  3  3 29
\end{verbatim}

There are two obvious ways to analyse this data set. First, analyse the meta-analysis of each imaging technology separately, which gives each study its own estimates of the hyperparameters. Second, analyse all studies together and incorporate the stratification using a technology-specific intercept.

To analyse all subdata separately, we call the function \code{meta4diag()} three times assuming for each subset model~\eqref{eq1} without covariate information. Here, we use the default settings
of \code{meta4diag()}.
\begin{verbatim}
R> res.CT = meta4diag(data = Scheidler[Scheidler$modality == "CT", ])
R> res.LAG = meta4diag(data = Scheidler[Scheidler$modality == "LAG", ])
R> res.MRI = meta4diag(data = Scheidler[Scheidler$modality == "MRI", ])
\end{verbatim}
Prior distributions as well as other model details, such as the link function, can be changed as described in Section~\ref{sec:data1}.

\begin{figure}[h!]
 \centering
 \subfloat[]{
    \centering
\includegraphics[width=0.5\textwidth]{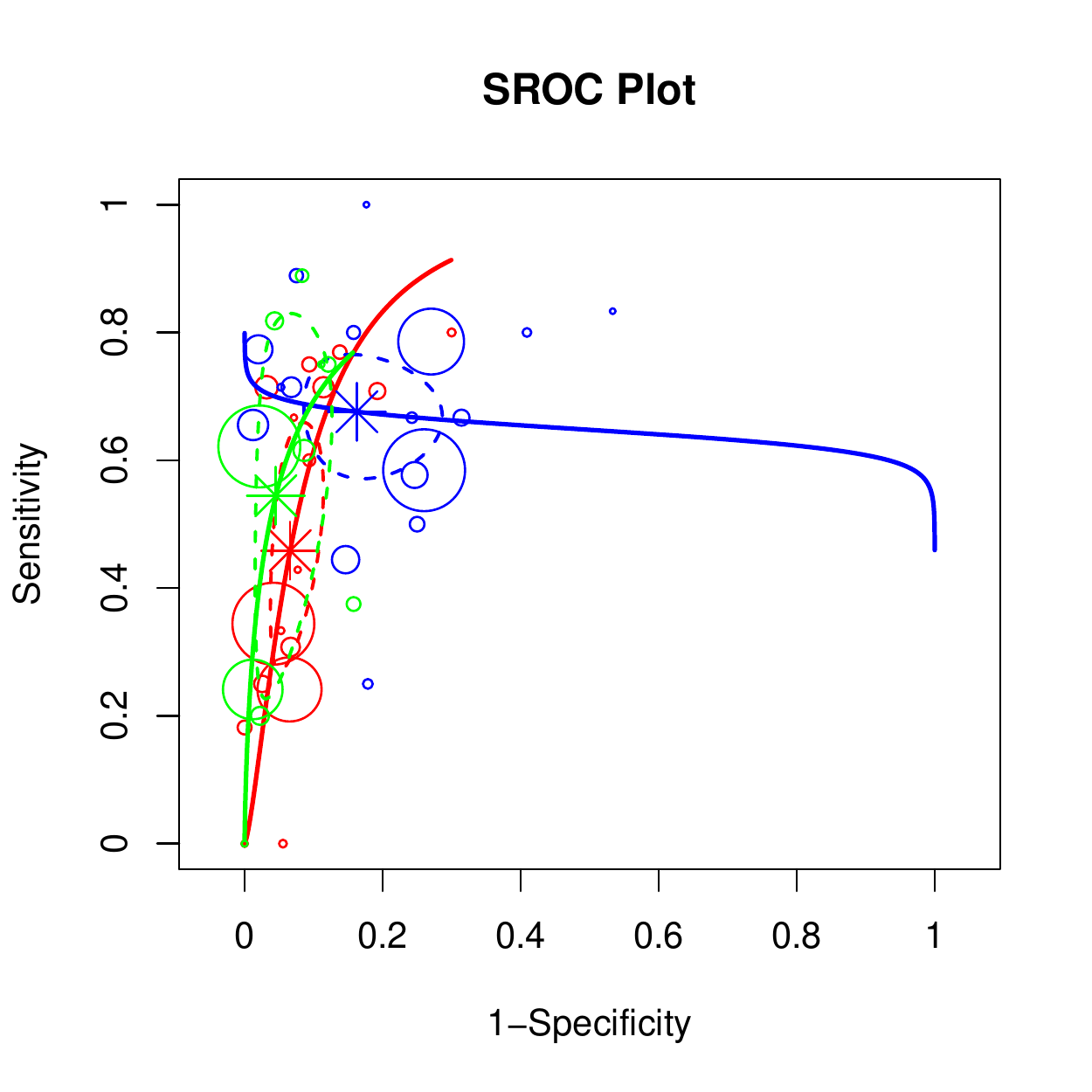}
    \label{fig:srocsub}
  }
  \subfloat[]{
    \centering
    \includegraphics[width=0.5\textwidth]{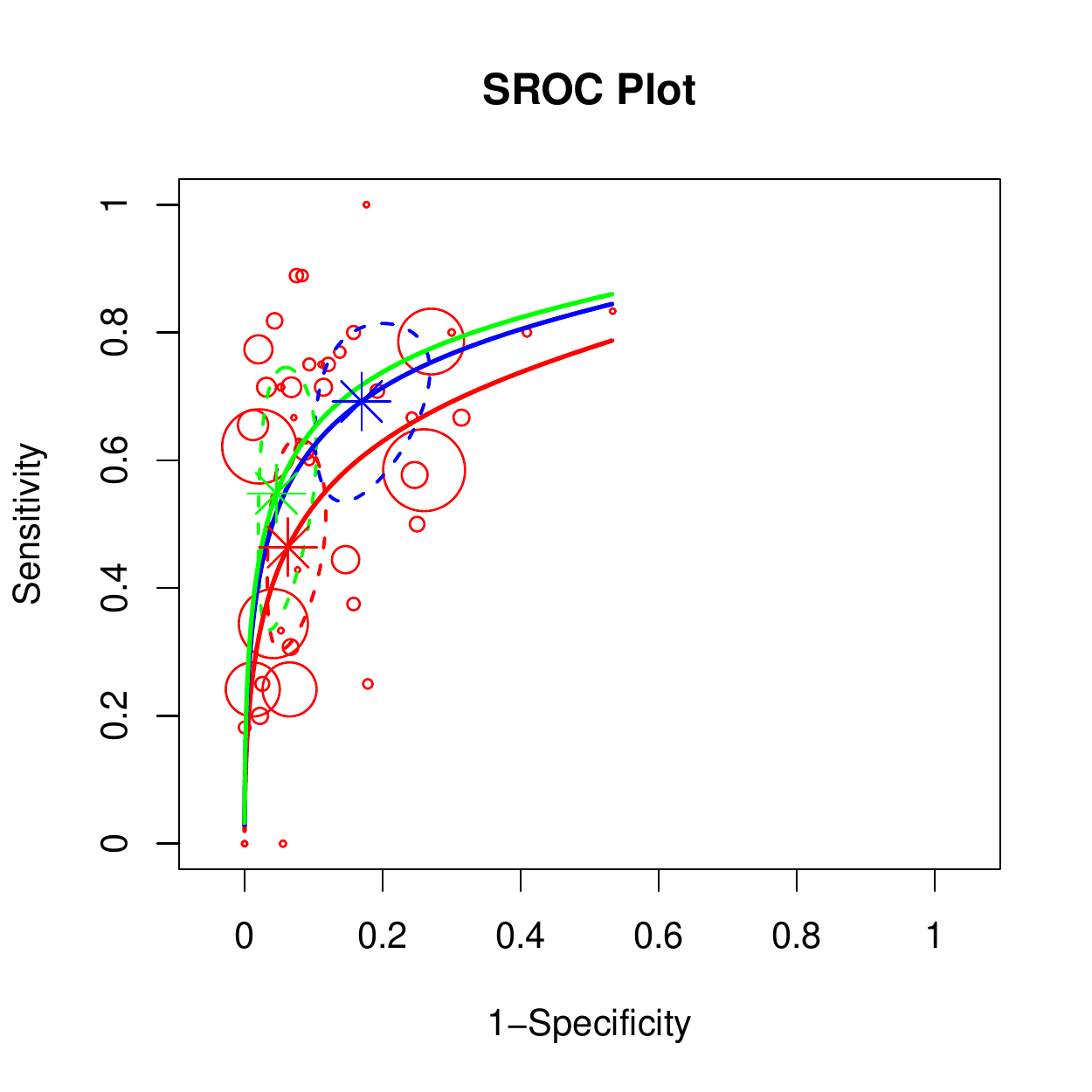}
    \label{fig:srocfull}
  }
	\caption{(a): SROC plot to compare the results for each sub-analysis of the \code{Scheidler} data set. Red: result for CT data. Blue: result for LAG data. Green: result for MRI data.
Bubbles represent the observed values where the size is proportional to the number of study participants, dashed lines are $95\%$ credible regions and the
star points are the summary points. The lines show the corresponding
SROC curves. (b): SROC plot for the joint analysis of the Scheidler
data set.  Bubbles are the observations, dashed lines are $95\%$
credible  regions and the star points are the summary points. The lines show the corresponding SROC curves. Red: result for CT data. Blue: result for LAG data. Green: result for MRI data.}
	\label{fig:compare}
\end{figure}

To plot the results of all three analyses in one device, we can use the \code{SROC()} function with the argument \code{add=TRUE}, see Figure~\ref{fig:srocsub}.
\begin{verbatim}
R> SROC(res.CT, dataShow = "o", lineShow = T, prShow = F, 
+    data.col = "red", cr.col = "red", sp.col = "red")
R> SROC(res.LAG, dataShow = "o", lineShow = T, prShow = F, 
+    data.col = "blue", cr.col = "blue", sp.col = "blue", add = T)
R> SROC(res.MRI, dataShow = "o", lineShow = T, prShow = F, 
+    data.col = "green", cr.col = "green", sp.col = "green", add = T)
\end{verbatim}
\setkeys{Gin}{width=\textwidth}
\begin{figure}[h!]
\centering
\end{figure}

To analyse the entire dataset, we consider the column ``modality" as a categorical covariate and use the following model where the overall intercept is omitted
\begin{equation}
\begin{aligned}
\label{eq:full}
&\textrm{TP}_i | \textrm{Se}_i\sim \textrm{Binomial}(\textrm{TP}_i+\textrm{FN}_i,\textrm{Se}_i),\quad \  \textrm{logit}(\textrm{Se}_i)=\mu_i+\phi_i,\\
&\textrm{TN}_i | \textrm{Sp}_i\sim \textrm{Binomial}(\textrm{TN}_i+\textrm{FP}_i,\textrm{Sp}_i),\quad \textrm{logit}(\textrm{Sp}_i)=\nu_i+\psi_i,\\
&\mu_i =  \left\{
             \begin{array}{lcl}
             \mu_{\textrm{CT}}  &\textrm{ if }& i=1,\dots,17\\
             \mu_{\textrm{LAG}} &\textrm{ if }& i=18,\dots,34\\
             \mu_{\textrm{MRI}} &\textrm{ if }& i=35,\dots,44
             \end{array}
        \right. \quad \quad
\nu_i =  \left\{
             \begin{array}{lcl}
             \nu_{\textrm{CT}} &\textrm{ if }& i=1,\dots,17\\
             \nu_{\textrm{LAG}} &\textrm{ if }& i=18,\dots,34\\
             \nu_{\textrm{MRI}} &\textrm{ if }& i=35,\dots,44
             \end{array}
        \right. \\
&\begin{pmatrix} \phi_i \\ \psi_i \end{pmatrix} \sim \mathcal{N}\left[ \begin{pmatrix} 0 \\ 0 \end{pmatrix},\begin{pmatrix} \sigma_{\phi}^2 & \rho\sigma_{\phi}\sigma_{\psi} \\ \rho\sigma_{\phi}\sigma_{\psi} & \sigma_{\psi}^2 \end{pmatrix}  \right].
\end{aligned}
\end{equation}

Here, $i = 1\dots 44$. To analyse this data in \code{meta4diag}, we call the function \code{meta4diag()} with argument \code{modality = "modality"}
\begin{verbatim}
R> res = meta4diag(data = Scheidler, modality = "modality")
R> res
\end{verbatim}
\begin{verbatim}
Time used: 
 Pre-processing    Running inla Post-processing           Total 
      0.1231420       0.4855158       0.1479652       0.7566230 

Model:Binomial-Normal Bivariate Model for Se & Sp. 
Data contains 44 primary studies. 

Data has Modality variable with level 3. 
Covariates not contained. 

Model using link function logit.

Marginals can be plotted with setting variable names to 
mu.CT, mu.LAG, mu.MRI, nu.CT, nu.LAG, nu.MRI, var1, var2 and rho. 
\end{verbatim}

To print the estimates for the parameters of the model, we use the function \code{summary()}
\begin{verbatim}
R> summary(res)
\end{verbatim}
\begin{verbatim}
Time used: 
 Pre-processing    Running inla Post-processing           Total 
      0.1231420       0.4855158       0.1479652       0.7566230 

Fixed effects: 
         mean    sd 0.025quant 0.5quant 0.975quant
mu.CT  -0.133 0.271     -0.675   -0.131      0.398
mu.LAG  0.775 0.263      0.262    0.773      1.300
mu.MRI  0.185 0.347     -0.504    0.186      0.869
nu.CT   2.598 0.268      2.074    2.596      3.133
nu.LAG  1.548 0.231      1.098    1.546      2.012
nu.MRI  2.927 0.342      2.255    2.926      3.607

Model hyperpar: 
           mean    sd 0.025quant 0.5quant 0.975quant
var_phi   0.800 0.308      0.353    0.747      1.551
var_psi   0.701 0.258      0.323    0.657      1.327
cor      -0.481 0.190     -0.790   -0.500     -0.058

-------------------
              mean    sd 0.025quant 0.5quant 0.975quant
mean(Se.CT)  0.467 0.057      0.359    0.467      0.575
mean(Se.LAG) 0.682 0.048      0.587    0.684      0.769
mean(Se.MRI) 0.545 0.072      0.405    0.546      0.678
mean(Sp.CT)  0.929 0.015      0.897    0.931      0.954
mean(Sp.LAG) 0.823 0.028      0.764    0.824      0.873
mean(Sp.MRI) 0.947 0.015      0.915    0.949      0.970

-------------------
Correlation between mu.CT and nu.CT is -0.3013. 
Correlation between mu.LAG and nu.LAG is -0.3494. 
Correlation between mu.MRI and nu.MRI is -0.3081. 

Marginal log-likelihood: -249.7198
Variable names for marginal plotting: 
      mu.CT, mu.LAG, mu.MRI, nu.CT, nu.LAG, nu.MRI, var1, var2, rho
\end{verbatim}

We apply the \code{SROC()} function again to check the difference between a separate and joint analysis
\begin{verbatim}
R> SROC(res, dataShow = "o", lineShow = T, prShow = F, 
+    cr.col = c("red", "blue", "green"), sp.col = c("red", "blue", "green"),
+    line.col = c("red", "blue", "green"))
\end{verbatim}
\setkeys{Gin}{width=\textwidth}
\begin{figure}[h!]
\centering
\end{figure}

Of note the SROC curves strongly vary depending on which formula is used to compute them, see \citet{chappell2009summary} for a discussion. Five different formulas
are available in \code{meta4diag} which can be chosen using the argument \code{sroc.type}, see Section~\ref{sec:data1} and documentation.

From Figure~\ref{fig:srocsub} and Figure~\ref{fig:srocfull}, we can see that the estimated summary points are almost the same in both analyses. However, the credible regions change slightly using the different model formulations. More striking are the changes in the SROC curves, in particular for the LAG subset (blue). Looking at the data there is no obvious trend that sensitivity increases along with increasing $1-$specificity. The estimated posterior correlation $\hat{\rho}$ is $0.1809 [-0.55, 0.79]$. \citet{chappell2009summary} stated that it is not appropriate to use SROC curves when $\hat{\rho}$ is close to zero or positive. Using separate analyses, we assume that each subdata has its own random effect properties. While using the full data set with a categorical covariate, we assume that all the subdata share the same covariance matrix.
The choice of how to model the data is up to the user.
However, when the argument \code{covariates} is used in the modelling, i.e., continuous covariates are included, the overall summary points, the confidence region and the prediction region are no longer available through the function \code{SROC()}, and only the study specified summary points can be obtained instead, see the example in Section~\ref{sec:data3}.

\begin{figure}[h!]
\centering
\includegraphics{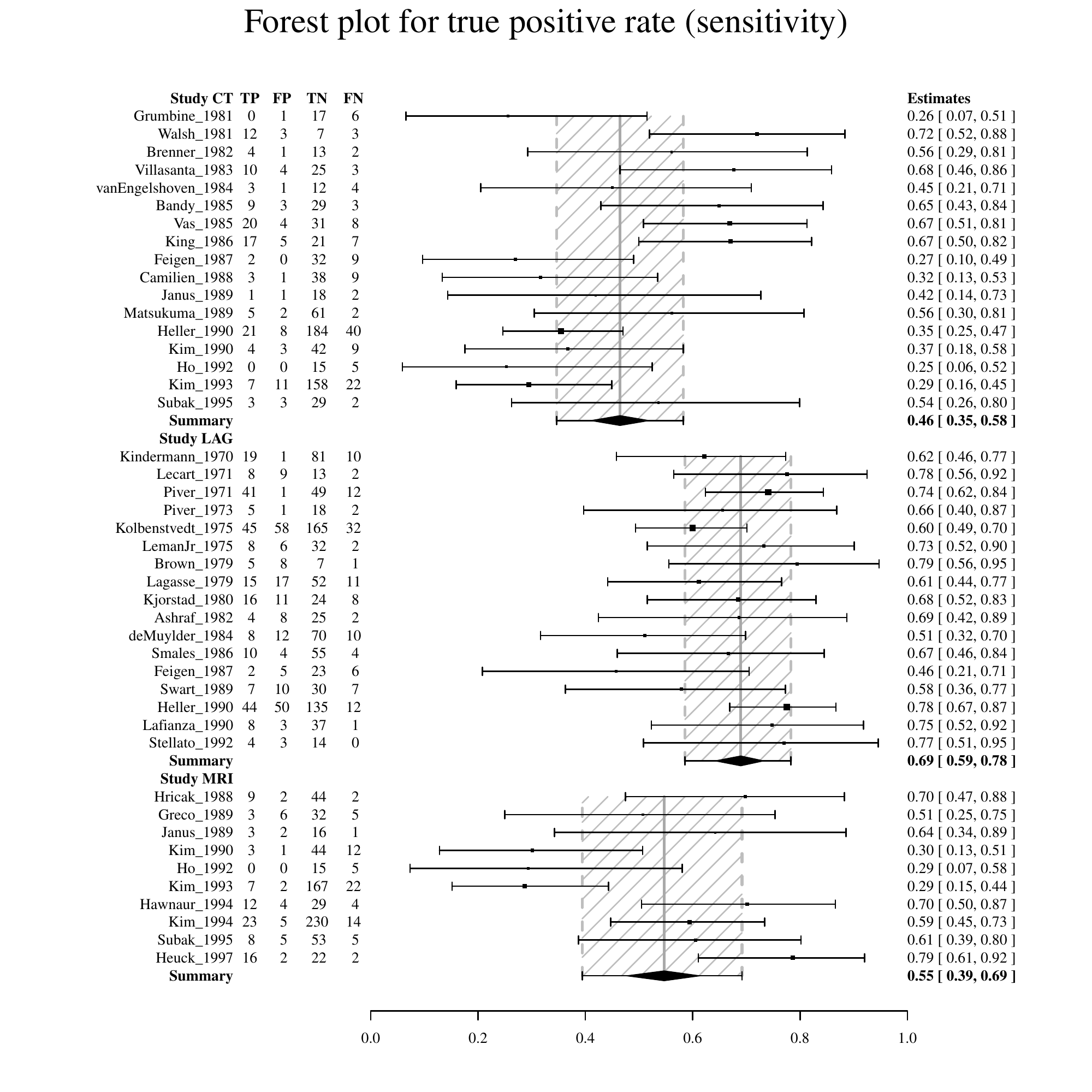}
\caption{Forest plot for the \code{Scheidler} data. The plot is separated into three parts relating to the three sub-data sets.\label{fig:forstfull}}
\end{figure}

The corresponding forest plot for this dataset is shown in Figure~\ref{fig:forstfull}. The plot is automatically separated into three parts due to the column \code{modality} with three different levels.
\begin{verbatim}
R> forest(res, accuracy.type = "sens")
\end{verbatim}
\setkeys{Gin}{width=\textwidth}
\subsection[Example 3]{Use of continuous covariate information} 
\label{sec:data3}

The Catheter Segment Culture data consists of $33$ studies from a diagnostic accuracy analysis by \citet{chu2009bivariate}. The studies analysed semi-quantitative (19 studies) and quantitative (14 studies) catheter segment culture for the diagnosis of intravascular device-related blood stream infection. In the dataset a column with name \code{type} indicates whether a study is based on semi-quantitative or quantitative catheter segment culture. We consider \code{type} as a categorical covariate in the model so that it should be set to \code{modality = "type"}. We choose this dataset as an example because it contains an additional column with name ``prevalence" providing disease prevalence information which can be considered as a continuous covariate. To analyse the dataset, we first load the \code{Catheter} dataset
\begin{verbatim}
R> data("Catheter", package = "meta4diag")
R> head(Catheter)
\end{verbatim}
\begin{verbatim}
      studynames              type prevalence TP  FP  TN FN
1    Cooper_1985 Semi-quantitative        3.6 12  29 289  0
2 Gutierrez_1992 Semi-quantitative       12.2 10  14  72  2
3 Cercenado_1990 Semi-quantitative       12.9 17  36  85  1
4     Rello_1991 Semi-quantitative       13.2 13  18  67  0
5      Maki_1977 Semi-quantitative        1.6  4  21 225  0
6 Aufwerber_1991 Semi-quantitative        3.1 15 122 403  2
\end{verbatim}

Consider that we would like to use the model
\begin{equation}
\begin{aligned}
\label{data2}
&\textrm{FN}_i | \textrm{Se}_i \sim \textrm{Binomial}(\textrm{TP}_i+\textrm{FN}_i,\textrm{Se}_i),\quad\quad\quad\quad\ \   \textrm{logit}(\textrm{Se}_i)=\mu_i + \alpha\cdot\textrm{prevalence}_i + \phi_i,\\
&\textrm{TP}_i | 1-\textrm{Sp}_i \sim \textrm{Binomial}(\textrm{TN}_i+\textrm{FP}_i,1-\textrm{Sp}_i),\quad \textrm{logit}(1-\textrm{Sp}_i)=\nu_i + \beta\cdot\textrm{prevalence}_i + \psi_i,\\
&\mu_i =  \left\{
             \begin{array}{lcl}
             \mu_{\textrm{semi-quantitative}}  &\textrm{ if }& i=1,\dots,19\\
             \mu_{\textrm{quantitative}} &\textrm{ if }& i=20,\dots,33
             \end{array}
        \right. \quad
 \nu_i =  \left\{
             \begin{array}{lcl}
             \nu_{\textrm{semi-quantitative}}  &\textrm{ if }& i=1,\dots,19\\
             \nu_{\textrm{quantitative}} &\textrm{ if }& i=20,\dots,33
             \end{array}
        \right. \\
&\begin{pmatrix} \phi_i \\ \psi_i \end{pmatrix} \sim \mathcal{N}\left[ \begin{pmatrix} 0 \\ 0 \end{pmatrix},\begin{pmatrix} \sigma_{\phi}^2 & \rho\sigma_{\phi}\sigma_{\psi} \\ \rho\sigma_{\phi}\sigma_{\psi}  & \sigma_{\psi}^2 \end{pmatrix}  \right].
 \end{aligned}
\end{equation}

That means we would like to model sensitivity and $1-$specificity jointly as proposed by \citet{chu2009bivariate} for this dataset. This can be done by setting \code{model.type = 2}. As the Catheter Segment Culture data contains one categorical covariate \code{type} and one continuous covariate \code{prevalence}, the argument \code{modality} is set to be \code{"type"} and argument \code{covariates} is set to be \code{"prevalence"}.
\begin{verbatim}
R> res = meta4diag(data = Catheter, model.type = 2, 
+    var.prior = "PC", var2.prior = "PC", cor.prior = "PC",
+    var.par = c(3, 0.05), cor.par = c(1, -0.1, 0.5, -0.95, 0.05, NA, NA),
+    modality = "type", covariates = "prevalence", 
+    quantiles = c(0.125, 0.875), nsample = 10000)
\end{verbatim}
Currently only one categorical covariate can be included in the model, whereas there is no limitation for the number of continuous covariates. In order to include more than one continuous covariate in the model, the user can provide a vector giving the names of all covariates to be included or the respective column numbers in the data frame. 

Here, we choose a PC prior for all hyperparameters. The vector of parameters for the PC prior of the correlation parameter must always be of length 7 specifying strategy, $\rho_0$, $\omega$, $u_1$, $\alpha_1$, $u_2$, $\alpha_2$. However, $u_2$ and $\alpha_2$ are not required when using \code{strategy = 1}, $u_1$ and $\alpha_1$ are not required when \code{strategy = 2} and there is no need to specify $\omega$ when \code{strategy = 3}, see Section~\ref{sec:prior}. To obtain the $12.5\%$ and $87.5\%$ quantiles in addition to the default $2.5\%$, $50\%$ and $97.5\%$ quantiles we set \code{quantiles = c(0.125, 0.875)}. Summary estimates are again obtained using the function \code{summary()}
\begin{verbatim}
R> summary(res)
\end{verbatim}
\begin{verbatim}
Time used: 
 Pre-processing    Running inla Post-processing           Total 
      0.4856298       0.7392230       0.1731811       1.3980339 

Fixed effects: 
                       mean    sd 0.025quant 0.125quant 0.5quant 0.875quant
mu.Semi.quantitative  1.692 0.349      1.032      1.302    1.681      2.089
mu.Quantitative       1.631 0.432      0.806      1.148    1.621      2.121
nu.Semi.quantitative -1.981 0.236     -2.450     -2.249   -1.981     -1.714
nu.Quantitative      -2.655 0.316     -3.288     -3.015   -2.652     -2.297
alpha.prevalence      0.007 0.015     -0.023     -0.010    0.007      0.023
beta.prevalence       0.032 0.012      0.008      0.019    0.032      0.046
                     0.975quant
mu.Semi.quantitative      2.415
mu.Quantitative           2.520
nu.Semi.quantitative     -1.517
nu.Quantitative          -2.038
alpha.prevalence          0.035
beta.prevalence           0.057

Model hyperpar: 
          mean    sd 0.025quant 0.125quant 0.5quant 0.875quant 0.975quant
var_phi  1.040 0.482      0.390      0.560    0.941      1.575      2.252
var_psi  0.764 0.232      0.415      0.521    0.726      1.031      1.321
cor      0.094 0.216     -0.326     -0.161    0.094      0.349      0.506

Marginal log-likelihood: -239.5213
Variable names for marginal plotting: 
      mu.Semi.quantitative, mu.Quantitative, nu.Semi.quantitative, 
      nu.Quantitative, alpha.prevalence, beta.prevalence, var1, var2, rho
\end{verbatim}

\begin{figure}[h!]
\centering
\includegraphics{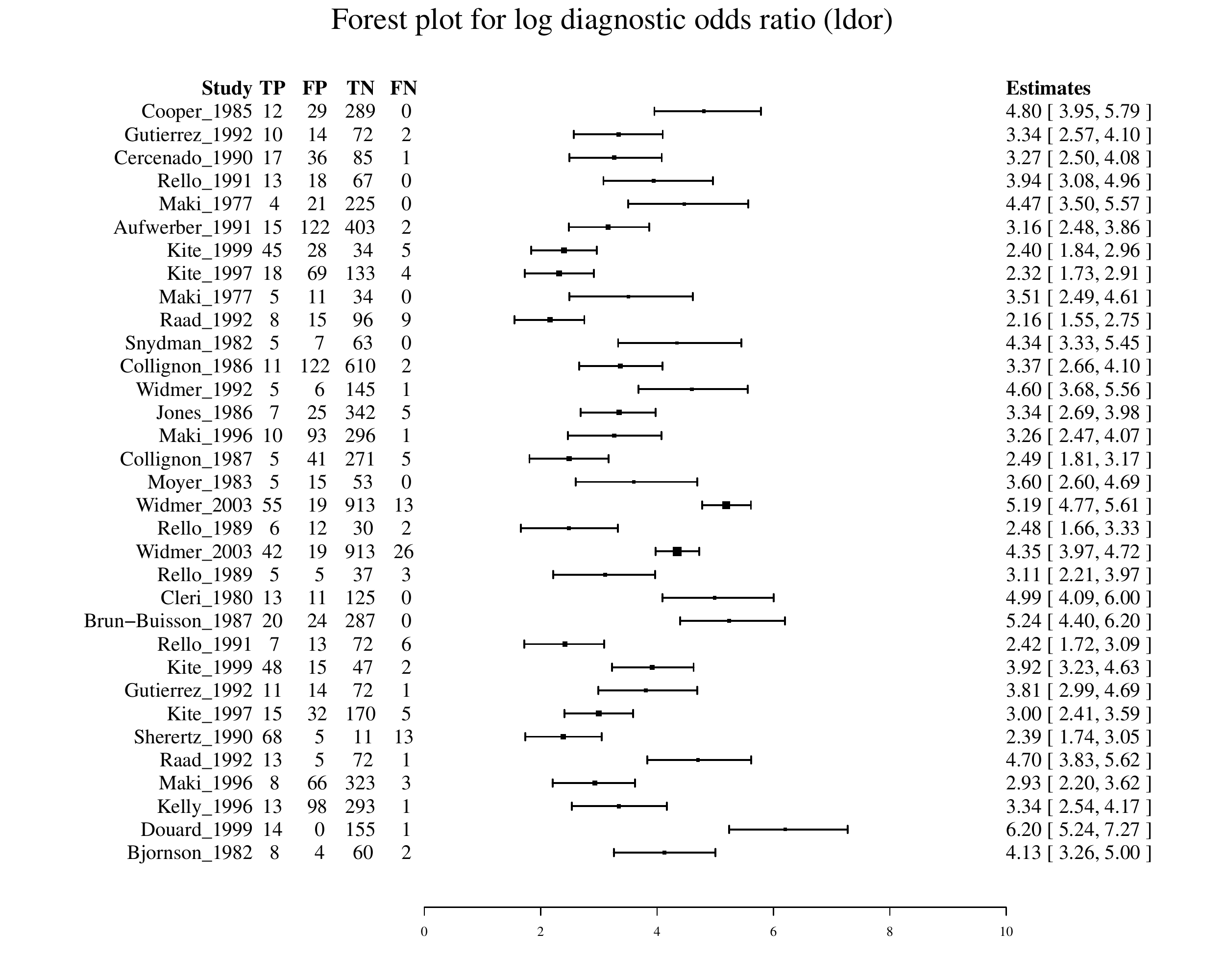}
\caption{Forest plot for the log diagnostic odds ratio (LDOR) of the \code{Catheter} data set. The study names, original dataset, estimated mean and corresponding $75\%$ credible intervals are also shown.}
\label{fig:forestplot2}
\end{figure}

A forest plot for the log diagnostic odds ratio is given in Figure~\ref{fig:forestplot2}. Here, $75\%$ credible intervals are shown which is specified by setting the argument \code{intervals = c(0.125, 0.875)} within the function \code{forest()}.
\begin{verbatim}
R> forest(res, accuracy.type = "LDOR", est.type = "median", 
+    nameShow = T, ciShow = "left", dataShow = "center", 
+    text.cex = 1.5, arrow.lwd = 1.5, 
+    cut = c(0, 10), intervals = c(0.125, 0.875))
\end{verbatim}
\setkeys{Gin}{width=\textwidth}

Of note, when the argument \code{covariates} is available, the summary estimates cannot be returned through the function \code{forest()}.
Similarly, the summary points, confidence region and prediction region in the SROC plot are not available. The SROC curve in contrast is still available. However, it does not depend on the choice of the argument \code{sroc.type}, but is computed according to \citet{walter2002properties} by fitting a regression equation
\begin{equation*}
D_i = a+bS_i
\end{equation*}
where
\begin{equation*}
D_i = \log\left(\frac{\widehat{\text{Se}}_i}{1-\widehat{\text{Se}}_i}\right)-\log\left(\frac{1-\widehat{\text{Sp}}_i}{\widehat{\text{Sp}}_i}\right)
\end{equation*}
and
\begin{equation*}
S_i = \log\left(\frac{\widehat{\text{Se}}_i}{1-\widehat{\text{Se}}_i}\right)+\log\left(\frac{1-\widehat{\text{Sp}}_i}{\widehat{\text{Sp}}_i}\right)
\end{equation*}
respectively. After fitting the regression line, the equation of the SROC curve can be obtained as
\begin{equation*}
\text{SROC}(x) = \frac{\exp(\frac{a}{1-b})x^{(1+b)/(1-b)}}{1+\exp(\frac{a}{1-b})x^{(1+b)/(1-b)}}, \quad x\in[0,1].
\end{equation*}
\subsection{Graphical user interface}
\begin{figure}[h!]
	\centering
	\includegraphics[width=\textwidth]{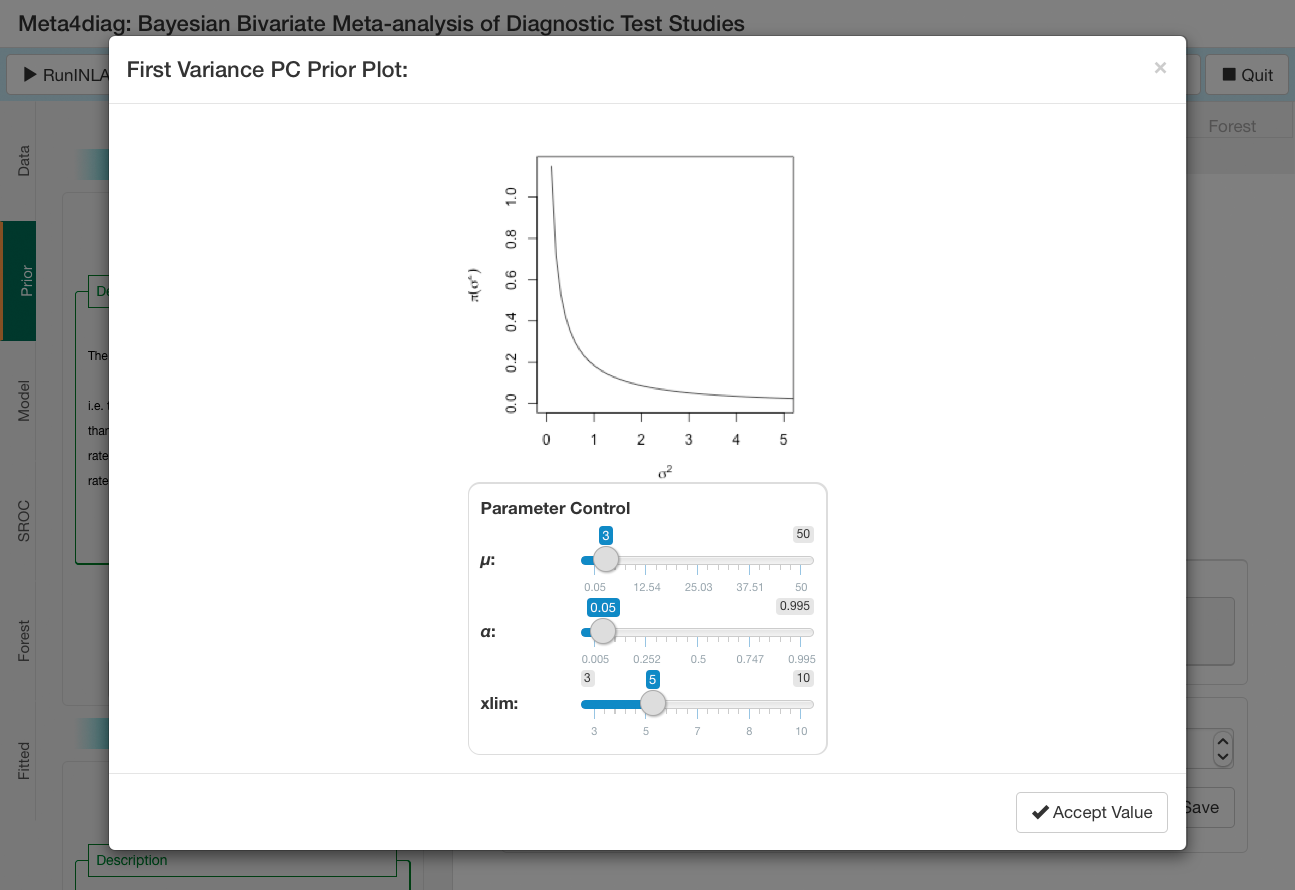}
	\caption{Interactive prior specification window. The prior density is shown and can be changed by sliding the bottom bars.}
	\label{guipriorchoose}
\end{figure}

To make Bayesian diagnostic meta-analysis easier to use for applied scientists, a cross-platform, interactive and user-friendly graphical user interface (GUI) has been implemented. The GUI can be used to load the data, set and graphically inspect the priors as the hyperparameters are manually changed by sliders (see Figure~\ref{guipriorchoose}), and run the model.
The results of the analysis  are shown directly in the interface and can be saved for later use.
The GUI only requires the basic knowledge of \proglang{R} required to start \proglang{R}, load the libraries and run the command that starts the GUI. Within the interface all options are visualised as buttons or drop-down menus, and
help for each option is found as tooltips when the user moves the mouse over the option or the ``Description area". The interface has been tested in the browsers ``Internet Explorer", ``Mozilla Firefox", ``Google Chrome" and ``Safari" on Linux, Mac and Windows 10 operating systems.

\begin{figure}[h!]
	\centering
	\includegraphics[width=\textwidth]{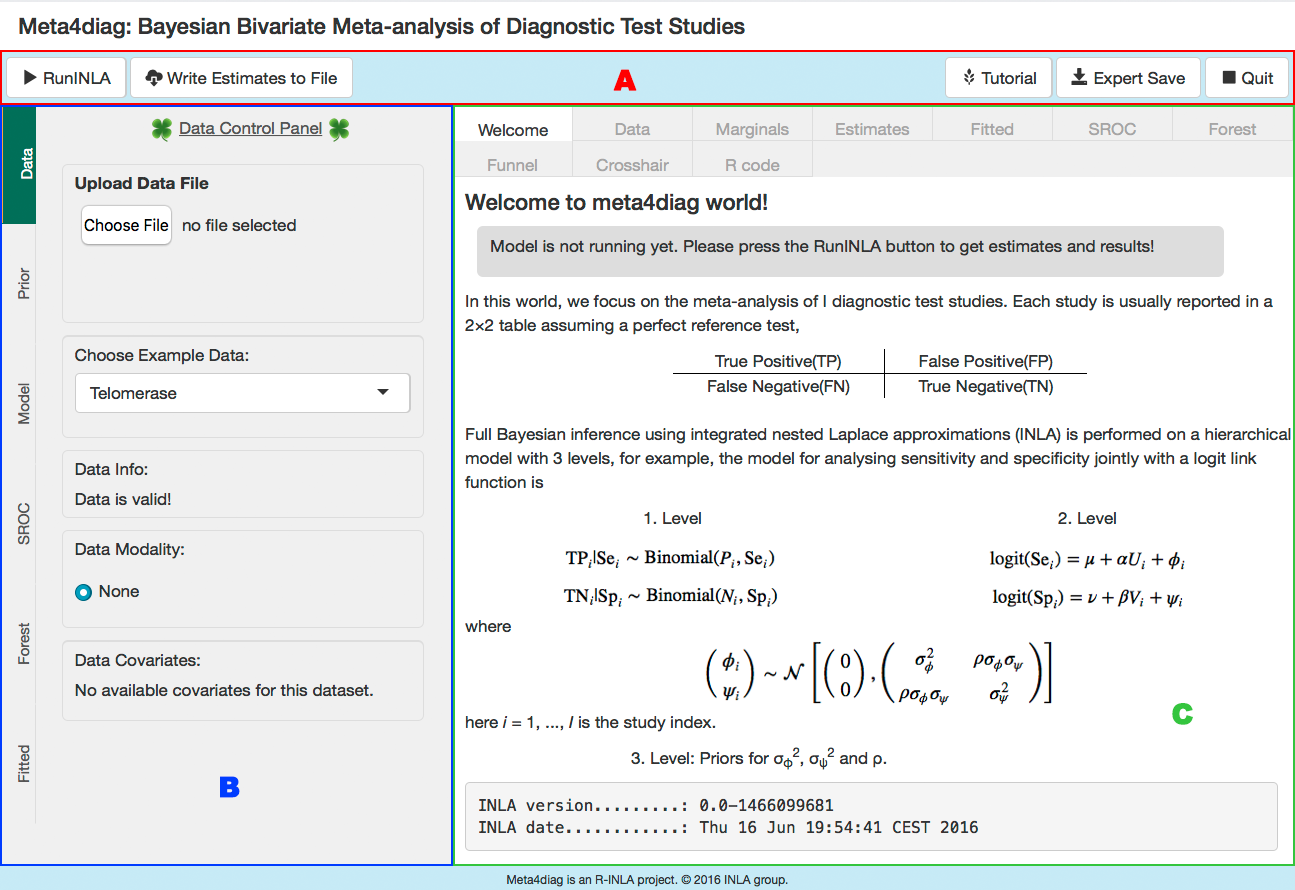}
	\caption{GUI main window of \pkg{meta4diag} after start up. (A) toolbar, (B) tool control panel, (C) view area, showing different pages (welcome message, data set, summary results, graphics). In particular, the ``Data Control Panel" is shown in the ``tool control panel" area. Users could upload their own datasets for analysing or choose a example dataset for understanding the package. The ``Welcome" page is shown in the ``view area". The basic information for modelling and the description of bivariate meta-analysis of diagnostic test studies are shown in this page.}
	\label{guistart}
\end{figure}

The GUI is started by loading the libraries \pkg{meta4diag}  and \pkg{INLA} and then calling the function \code{meta4diagGUI()} with
\begin{verbatim}
R> library("meta4diag")
R> library("INLA")
R> meta4diagGUI()
\end{verbatim}
The start window of the GUI is shown in Figure~\ref{guistart}  and is divided into three areas A, B and C.
A contains the toolbar and has buttons for running INLA and writing the results to a text file, and buttons for starting the tutorial, saving the results to an \proglang{R} object
for further study in \proglang{R} and for quitting the interface. B has 6 tabs that contain the various control panels, which are used to set up the analysis, such as the data control
panel, the prior control panel, and the model control panel. The options within these three panels must be set  before pressing the
``RunINLA" button in A. The ``Forest" control panel and ``SROC" control panel in B  are used for choosing plotting settings, and can be used both before and after running
the model, and the ``Fitted" panel allows the user to inspect the estimates for different choices of accuracy types and can also be set after running the
model. Lastly, C has 10 tabs where the first is a welcome page and the rest are used to view the data and the results.

\begin{figure}[h!]
	\begin{minipage}[t]{0.5\linewidth}
	\centering
	\includegraphics[width=\textwidth]{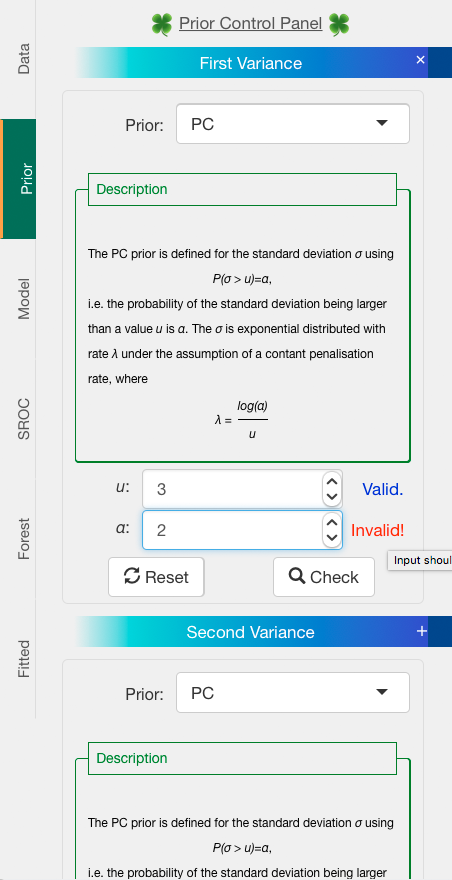}
	\end{minipage}
	\begin{minipage}[t]{0.5\linewidth}
	\centering
	\includegraphics[width=\textwidth]{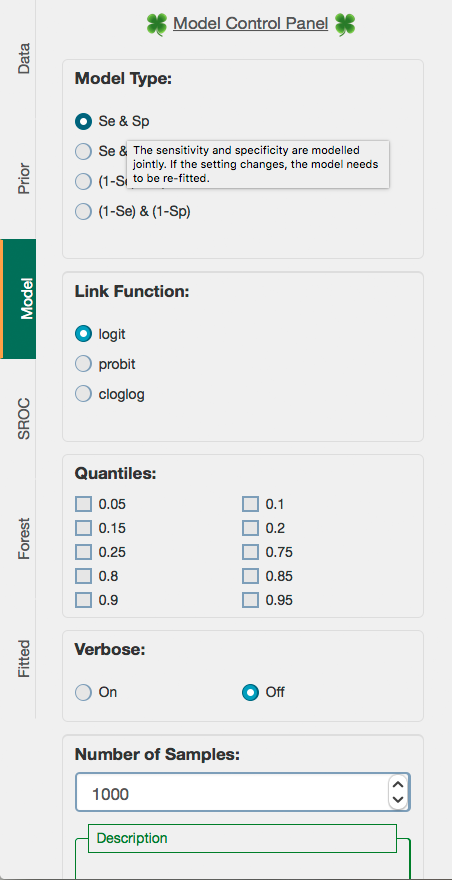}
	\end{minipage}
	\caption{Details for two tool panels. Left: ``Prior Control Panel". In this panel, users can set the prior distributions for the first variance component, second variance component and the correlation. In particular, the specification of the PC-prior for the first variance component is shown. A ``Description area" is shown to explain what the prior is. The red ``Invalid!" indicates that the given value for the hyper parameter $\alpha$ is not valid. The interval of the valid values can be seen from the tooltips of the indicator ``Invalid".  Right: ``Model Control Panel".}
	\label{guipanel}
\end{figure}

Figure~\ref{guipanel}, which contains a screenshot of  the ``Prior Control Panel" on the
left-hand side and the ``Model
Control Panel" on the right-hand side, gives an example of how the user can set the model and the prior.
The description of the options in each panel is integrated in the GUI through tooltips, but can also  be found in the package
documentation (see
\code{meta4diag()} for details). The left-hand side screenshot only shows how the user can
set the prior distribution for the first
variance component, but the panel also contains options for setting the priors on the
second variance component and the correlation in
the bivariate model. Figure~\ref{guipriorchoose} illustrates how the
user can explore different settings of the hyperparameters interactively by
sliding the sliders corresponding to each parameter. When the PC prior is selected
for the correlation parameter, the user may use either of the specification
strategies described in Section~\ref{sec:prior}
 to set the hyperparameters.
The ``Model
Control Panel"  shown on the right-hand side of Figure~\ref{guipanel} is used to specify the model
type, link function, quantiles of interest, and more.

\begin{figure}[h!]
	\centering
	\includegraphics[width=\textwidth]{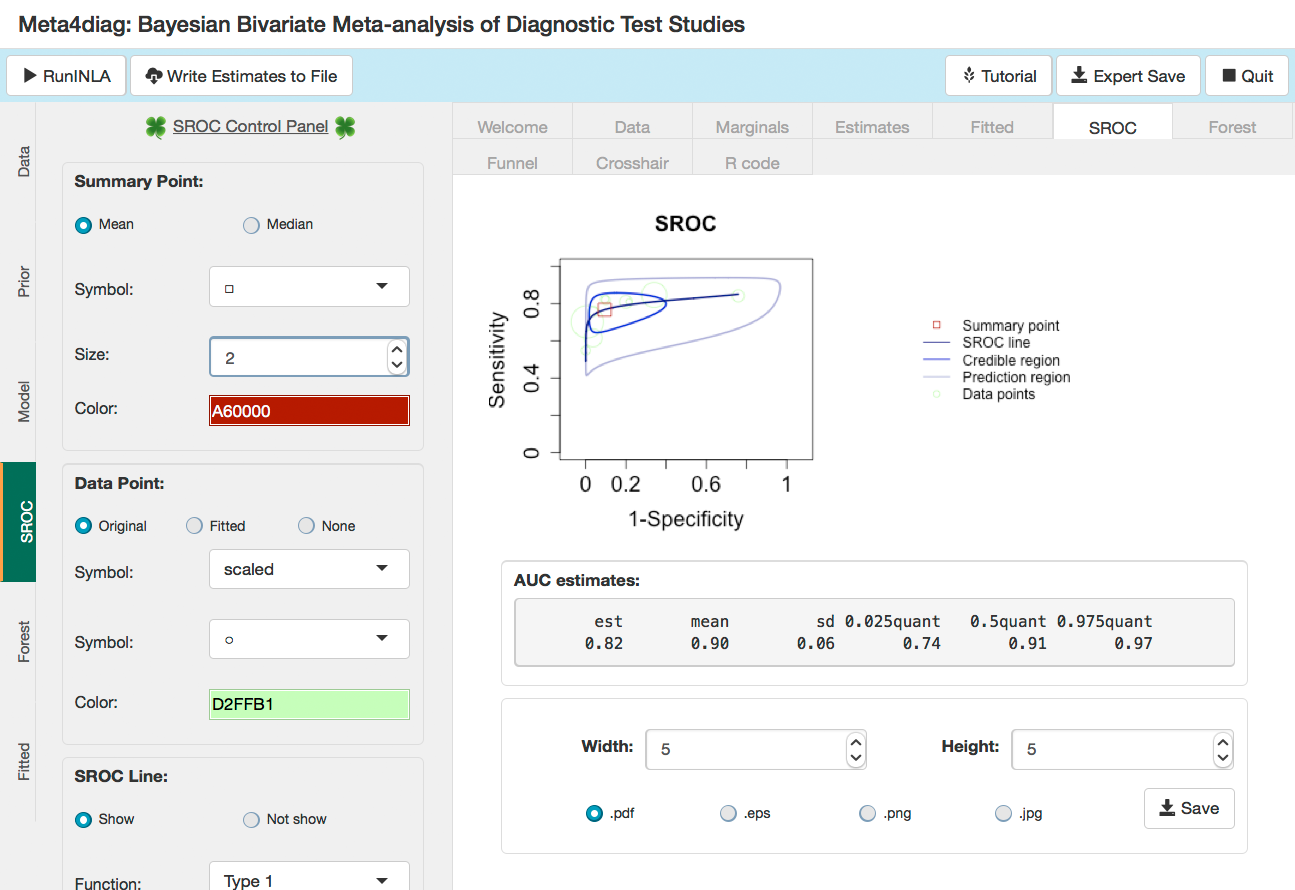}
	\caption{Example of an SROC plot in the
          view area and the ``SROC control panel" in the left tool bar.}
	\label{guiresult}
\end{figure}

After setting the options in the first three control panels and clicking the ``RunINLA"
button, the chosen dataset will be loaded and analysed. The results of the analysis
will be shown in the view area (C) and, for example, the SROC plot can be viewed in the SROC tab
of the view area (C) as shown in Figure~\ref{guiresult}. The other tabs can be used to view summary
estimates, study-specific accuracy estimates, posterior marginal plots
and forest plot, and in each case  the \proglang{R} code that generated the figure or text is also shown.
If the data, the model or prior settings are changed, the user must push
the ``RunINLA'' button again to update the results.

\section[Conclusion]{Conclusion}\label{sec:conc}
The present paper demonstrates the usage of the novel R-package \pkg{meta4diag} for analysing bivariate meta-analyses of diagnostic test studies with \proglang{R}, and illustrates its usage using three examples from the literature. The package is built on top of the R-package \pkg{INLA} and thus provides full Bayesian inference without the need for Markov Chain Monte Carlo techniques. This is especially important when several or complex meta-analyses are studied, or simulation studies shall be performed, as then the time speed-up becomes obvious. The model can be easily specified, whereby the user does not need to know any \pkg{INLA}-specific details. Quantities relevant in the field of diagnostic meta-regression are internally computed and returned directly without requiring the user to work with the general and complex \pkg{INLA} output.

One of the biggest advantages, besides of being fast, compared to
other software packages for Bayesian inference is the flexible and at the same time intuitive
prior specification framework. In particular the newly proposed PC
priors \citep{2014arXiv1403.4630S} are supported. Here, the user has
the possibility to incorporate expert knowledge in the form of
probability contrasts. \citet{2015arXiv151206217G} compared the
performance of different PC priors with previous proposed priors in
the bivariate model through an intensive simulation study and a real
data set. Both informative and less informative PC priors were
studied, and results indicated that the PC priors perform at least as
good as previously used priors.

A graphical user interface makes the package also attractive for users who prefer to work with interactive windows offering selection menus. The GUI provides the full functionality of the package. In addition the user can inspect the priors directly and change them interactively. By offering fast inference within a Bayesian framework, intuitive choice of prior distributions and the GUI we feel that this package has great potential for routine practice. As a future research direction, we would like to expand the functionality of this package to a three-variate model analysing sensitivity, specificity and disease prevalence jointly. Further, we would like to investigate how to extend \pkg{meta4diag} when the assumption of a perfect reference test is not given.

\section[Acknowledgments]{Acknowledgments}

We are grateful for helpful discussions with H{\aa}vard Rue and Geir-Arne Fuglstad.

\bibliography{meta4diagBIB}
\end{document}